\documentclass[preprint2]{aastex}

\usepackage{graphicx,caption,subcaption,color}
\newcommand{\tc}{\textsl{The~Cannon}} 
\newcommand{\apogee}{{APOGEE}}

\newcommand{\aspcap}{\textsl{ASPCAP}}

\newcommand{\badstar}{\texttt{STAR BAD}} 
\newcommand{\aspcapflag}{\texttt{ASPCAPFLAG}}

\newcommand{\teff}{\mbox{$\rm T_{eff}$}}
\newcommand{\kms}{\mbox{$\rm kms^{-1}$}}
\newcommand{\feh}{\mbox{$\rm [Fe/H]$}}

\newcommand{\alphafe}{\mbox{$\rm [\alpha/Fe]$}}

\newcommand{\logg}{\mbox{$\rm \log g$}}

\newcommand{\rgc}{\mbox{$\rm R_{GC}$}}
\newcommand{\vgal}{\mbox{$\rm V_{GAL}$}}

\begin{document}

\title{APOGEE Kinematics I: Overview of the Kinematics of the Galactic Bulge as Mapped by APOGEE}
\author{M.~Ness\altaffilmark{1},  
G.~Zasowski\altaffilmark{2}, 
J.A.~Johnson\altaffilmark{3}, 
E.~Athanassoula\altaffilmark{4}, 
S.R.~Majewski\altaffilmark{5}, 
A.E.~Garc\'{i}a~P\'{e}rez\altaffilmark{6,7},
J.~Bird\altaffilmark{8}, 
D.~Nidever\altaffilmark{9}, 
Donald P. Schneider\altaffilmark{10,11},
J.~Sobeck\altaffilmark{5},
P.~Frinchaboy\altaffilmark{12},
Kaike Pan\altaffilmark{13},
Dmitry Bizyaev\altaffilmark{13,14},
Daniel Oravetz\altaffilmark{13},
Audrey Simmons\altaffilmark{13}}
\altaffiltext{1}{Max-Planck-Institut f\"ur Astronomie, K\"onigstuhl 17, D-69117 Heidelberg, Germany}
\altaffiltext{2}{Department of Physics \& Astronomy, Johns Hopkins University, Baltimore, MD, 21218, USA}
\altaffiltext{3}{Department of Astronomy and Center for Cosmology and AstroParticle Physics (CCAPP), Columbus, OH 43210, USA  }
\altaffiltext{4}{Aix Marseille Universit\'e, CNRS, LAM (Laboratoire d'Astrophysique de Marseille) UMR 7326, 13388, Marseille, France }
\altaffiltext{5}{Department of Astronomy, University of Virginia, Charlottesville, VA 22904-4325, USA}
\altaffiltext{6}{Instituto de Astrof\'{\i}sica de Canarias, 38205 La Laguna, Tenerife, Spain}
\altaffiltext{7}{Departamento de Astrof\'{\i}sica, Universidad de La Laguna, 38206 La Laguna, Tenerife, Spain}
\altaffiltext{8}{Department of Physics and Astronomy, Vanderbilt University,    6301 Stevenson Center, Nashville, TN, 37235}
\altaffiltext{9}{Department of Astronomy, University of Michigan, Ann Arbor, MI 48104, USA} 
\altaffiltext{10}{Department of Astronomy and Astrophysics, The Pennsylvania State University, University Park, PA 16802}
\altaffiltext{11}{ Institute for Gravitation and the Cosmos, The Pennsylvania State University,   University Park, PA 16802}
\altaffiltext{12}{Department of Physics \& Astronomy, Texas Christian University, TCU Box 298840, Fort Worth, TX 76129, USA}
\altaffiltext{13}{Apache Point Observatory and New Mexico State University, P.O. Box 59, Sunspot, NM, 88349-0059, USA}
\altaffiltext{14}{Sternberg Astronomical Institute, Moscow State University, Moscow}

\email{ness@mpia.de}

\begin{abstract}%

We present the stellar kinematics across the Galactic bulge and into the disk at positive longitudes from the SDSS-III \apogee\ spectroscopic survey of the Milky Way. \apogee\ includes extensive coverage of the stellar populations of the bulge along the mid-plane and near-plane regions. From these data, we have produced kinematic maps of 10,000 stars across longitudes 0$^\circ$ $<$ $l$ $<$ 65$^\circ$, and primarily across latitudes of $|b|$ $<$ 5$^\circ$ in the bulge region.  The APOGEE data reveal that the bulge is cylindrically rotating across all latitudes and is kinematically hottest at the very centre of the bulge, with the smallest gradients in both kinematic and chemical space inside the inner-most region $(|l,b|)$ $<$ (5$^\circ$,5$^\circ$). The results from APOGEE show good agreement with data from other surveys at higher latitudes and a remarkable similarity to the rotation and dispersion maps of barred galaxies viewed edge on. The thin bar that is reported to be present in the inner disk within a narrow latitude range of $|b|$ $<$ 2$^\circ$ appears to have a corresponding signature in \feh\ and \alphafe.  Stars with \feh\ $>$ --0.5 have dispersion and rotation profiles that are similar to that of N-body models of boxy/peanut bulges. There is a smooth kinematic transition from the thin bar and boxy bulge $(l,|b|)$ $<$ (15$^\circ$, 12$^\circ$) out into the disk for stars with \feh\ $>$ -1.0, and the chemodynamics across $(l,b)$ suggests the stars in the inner Galaxy with \feh\ $>$ -1.0 have an origin in the disk. 

\end{abstract}


\section{Introduction}\label{sec:Intro}

The Galactic bulge of the Milky Way is a cylindrically rotating system \citep[e.g.,][]{howard2008, kunder2012, zoccali2014} for the stars with ${\rm [Fe/H]} > -1.0$ \citep{Ness2013b}. These ${\rm [Fe/H]} > -1.0$ stars, which comprise the majority of stars in the bulge \citep[e.g.,][]{zoccali2008}, have a smooth kinematic transition out into the disk, as well as symmetry about the major axis. Previous studies of the bulge, which have mostly examined stars at $|b| > 4^\circ$, have demonstrated that there is a kinematic dependence of the stars in the bulge on [Fe/H]. The stars with ${\rm [Fe/H]} > -0.5$, which are part of the X-shaped profile of the Milky Way \citep{Ness2012, Uttenthaler2012}, show a velocity dispersion profile that is consistent with bulge formation via dynamical instabilities seen in N-body models. The most metal-rich subset of these stars (i.e., \feh\ $>$ 0) are kinematically colder than their more metal-poor counterparts ($-0.5 < {\rm [Fe/H]} < 0$), with the same characteristic dispersion profile, as a function of $(l,b)$. This dependence of the kinematic profile on metallicity for stars that appear to be part of the bulge/bar is not yet fully understood, but is likely related to the mapping of the stars into the boxy bulge from the disk by the bar, as a function of their initial phase-space in the disk \citep{Inma2013, PdiM2015}. \\

Bulge stars in the yet lower metallicity range, {of} $-1.0 < \feh\ < -0.5$ show a similar rotation, but dissimilar dispersion, to the more metal-rich stars. The dispersion profile of these more metal-poor stars does not display a strong dependence on latitude and appears dissimilar to that of N-body models of bulge formation via dynamical instabilities of the bar.  These stars are possibly part of the thick disk population in the inner Milky Way, given their fast rotation and high, latitude-independent dispersion \citep[e.g.,][]{Ness2013b, PdiM2015}. Alternatively, these stars may be part of a unique, classical bulge component of the Milky Way \citep[e.g.,][]{Hill2011, Babusiaux2010}. It has recently also been suggested that these stars could also have an origin in massive clumps at higher redshift \citep{zoccali2014}. {The RR Lyrae stars which peak at a metallicity of \feh\ $<$ --1.0 in the bulge follow an ellipsoidal density distribution along the bar \citep{Piet2014}. The most metal-poor stars, with $\feh\ < -1.0$, comprise about 5\% of stars in the bulge and these are slowly rotating; they may be halo stars in the inner region or possibly part of a population that is unique to the bulge formed at the earliest epoch of Galaxy assembly \citep{diemand2008, Casey2015, Howes2015}. }

 The bulge fields of the \apogee\ survey \citep{Majewski2012, Majewski2015}, are observed in the infrared and thus are uniquely and critically placed to characterize previously unexamined regions of the bulge. {Many previous bulge studies have revealed the kinematic trends for fields in the bulge located off the plane ($|b| > 4^\circ$)}. {Additionally, the GIBS survey  \citep{zoccali2014} has examined 14 fields within $(l,b)$ $<$ $(10^\circ,4^\circ)$, as part of their combined 24 field sample of 5000 red clump giants and 1200 red clump stars toward the bulge}.  Using the \apogee\ data, which trace unexamined regions across the bulge into the disk, including at $b$ $<$ 4$^\circ$, we address the following questions: (1) What is the nature of the kinematic profile of the bulge in previously unexplored regions in and near the plane? (2) Are the kinematics in these previously unexplored regions at odds or consistent with the kinematics measured off the plane and with N-body models? (3) What is the dependence of the kinematic profile of the stars on \feh? % This is key to reconstructing the formation history of the bulge.% from archive data in the bulge, observed at resolution R = 6500 (plus 450 stars in a field at $b$  = --3.5$^\circ$ observed at a high resolution of R = 22,500 (the same resolution as the \apogee\ survey)

 The high resolution, $R = 22,500$ spectra spanning 1500 - 1700 nm allow not only the first homogeneous, contiguous mapping of the bulge metallicity and kinematics through the plane of the Milky Way into the inner-most bulge region, but also a mapping of the kinematics as a function of multi-element abundances.  For the characterisation of stellar populations in these previously unexplored regions presented in this paper, we use the APOGEE spectra from the SDSS-III Data Release 12 \citep[DR12;][]{Ahn2014} for about 20,000 stars toward the Galactic Bulge and surrounding disk (Section~\ref{sec:params_and_distances}). These stars span mostly positive longitudes, from $-5^\circ < l < 65^\circ$, across $|b|$ $<$ 5$^\circ$ and including some higher latitude fields at $b$ $<$ 15$^\circ$.  We have implemented distance cuts in order to select the 10,000 stars that are located at distances 4-12 kpc along the line of sight.

\section{Stellar Parameters and Distances} \label{sec:params_and_distances}

The APOGEE survey, part of the SDSS-III project \citep[][]{Eisenstein2011},
operates at the 2.5-meter telescope of the Apache Point Observatory \citep{Gunn2006}.
Being a near-IR survey, APOGEE has observed a large, homogeneous sample of stars ($\sim$150,000) spanning the halo,
disk, and innermost regions of the Milky Way, including the boxy bulge \citep[for details of the target selection, see][]{Zasowski2013}.  All data are processed with a custom pipeline \citep{Nidever2015}, 
which also calculates the stellar radial velocities with an accuracy of $<1$~km~s$^{-1}$.

{ We use the publicly available spectra, released as part of DR12, to determine stellar parameters of \teff, \logg, \feh\ and \alphafe\ using \tc\ \citep{Ness2015}. The set of reference objects for training \tc\ is comprised of a subsample of $\approx$ 2150 high signal to noise (SNR $\sim$ 300) \apogee\ stars with no \badstar\ set in \aspcapflag\  as well as the globular and open cluster stars \citep{Meszaros2013}. We adopt the high signal-to-noise data and their measurements from \aspcap\ as our reference objects in order to determine precise stellar parameters for the lower signal to noise \apogee\ data, as described in \citet{Ness2015}. We use the corrected \teff, \logg, \feh\ and \alphafe\ labels from \aspcap\ \citep{GP2015}  for these training stars (the corrections to these parameters is done by \apogee\ using the cluster calibration sample \citep{Meszaros2013}). We also include the metal-poor ($\feh\ < -1.0$) globular cluster stars with their isochrone-corrected training labels \citep{Ness2015} so as to extend our training set to the lowest metallicities. The Pleiades open cluster, which contains the main sequence stars in the training set, enables dwarf stars in the DR12 data to be identified by \tc\ and therefore eliminated.  In total there are 2400 reference objects that are used for the training step of \tc. The \alphafe\ from the \apogee\ spectra is comprised of contributions from many alpha-elements including O, Ca, Ti, Si and Mg \citep[see][]{GP2015}.}

\begin{figure}[h]
\centering
\vspace{-0pt}
        \includegraphics[scale=0.28]{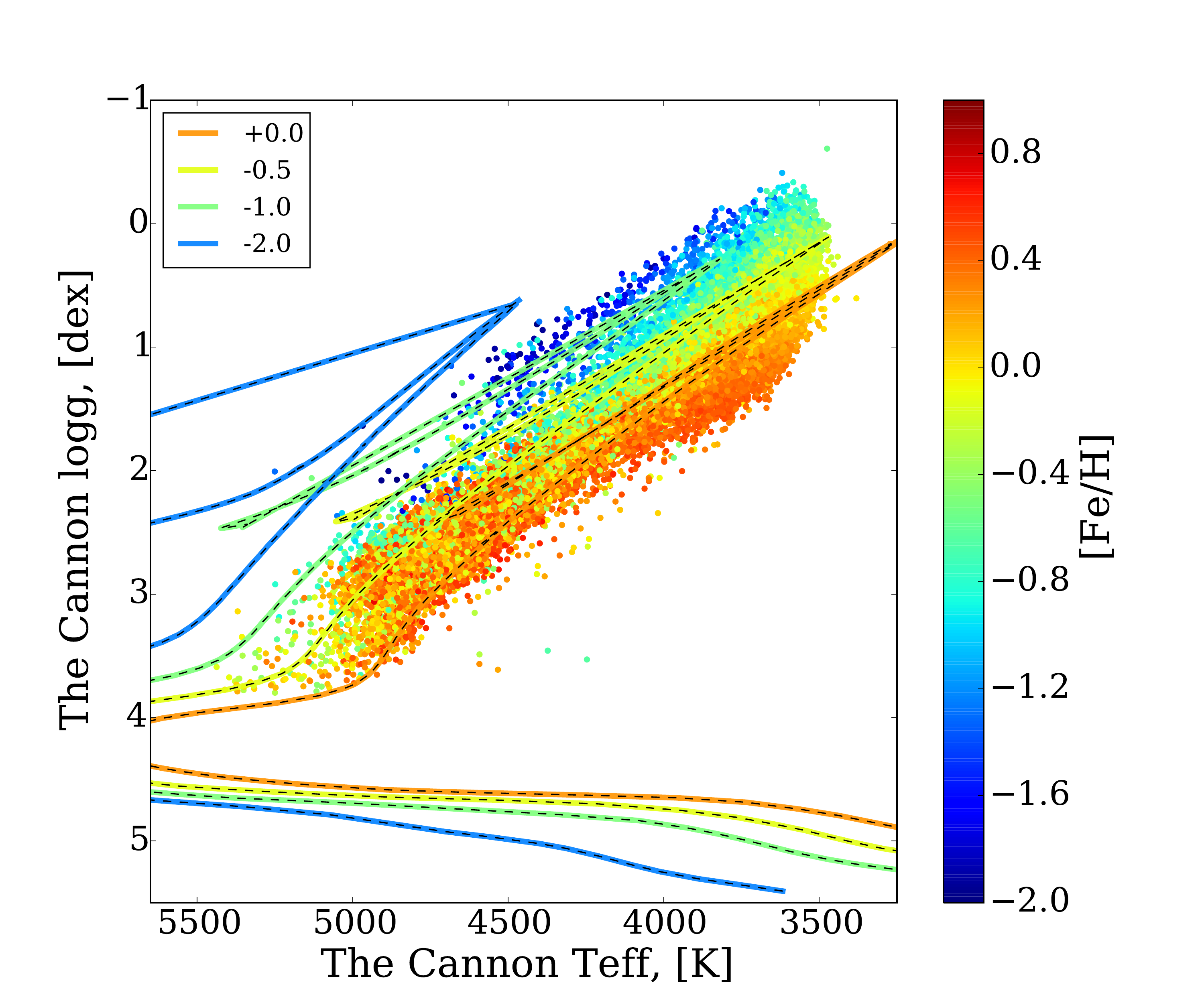}\\
\caption{{The \teff-\logg\ plane for the $\approx$ 20,000 stars selected for the analysis of the bulge and inner disk, located in fields spanning $-5^\circ < l < 65^\circ$,  $b$ $<$ 15$^\circ$.  Stellar parameters have been determined with \tc\ \citep{Ness2015}. These stars are the full sample for which we have implemented distance cuts in order to select the 10,000 stars that are located at distances 4-12 kpc along the line of sight.  PARSEC isochrones of 10 Gyr \citep{Bressan2012} are shown and the metallicity color bar of the stars is the same as the plotted isochrones, the metallicity index for the isochrone is shown in the top left hand corner.}}
\vspace{-0pt}
\label{fig:stars}
\end{figure}

For our analysis, standard and telluric stars were removed from the sample, as well as flagged rapid rotators and special targets from ancillary programs. The main survey objects were selected using the \apogee\ \textit{EXTRATARG} flag. Only giants were chosen for the analysis with \logg\ $<$ 3.8. Following this selection, there remain $\approx$ 20,000 stars in the direction of the bulge and inner disk, across $|l|$ $<$ 65$^\circ$ and $|b|$ $<$ 15$^\circ$. These $\approx$ 20,000 stars are shown in Figure~\ref{fig:stars} (no distance cuts have yet been applied to this sample). We note these have had no corrections or calibrations applied to them, nor prior information about the \teff-\logg\ space of the isochrones; these results are directly out of \tc's label-estimation step. 

{The parameter range of \tc's training set is narrower than that of the \apogee\ stars toward the bulge and disk. The training set spans ranges of 3600~K $<$ \teff\ $<$ 5750~K,  -0.3 $<$ \logg\ $<$ 4.7 , -2.5 $<$ \feh\ $<$ 0.45, and -0.3 $<$ \alphafe\ $<$ 0.5.  The range of parameters of the giant stars toward the bulge and disk determined by \tc\ for the survey spectra is 3500~K $<$ \teff\ $<$ 5500~K,  --0.5 $<$ \logg\ $<$ 3.8, --2.0 $<$ \feh\ $<$ 0.7 and --0.3 $<$ \alphafe\ $<$~0.5. \apogee\ stars that have stellar parameters determined by \tc\ outside of that of the training data, are assigned values based on extrapolation of \tc's model. Therefore the scale of these parameters is not tied directly to any physical calibration. }

{Nevertheless, as seen in Figure \ref{fig:stars}, almost all stars fall naturally on the isochrone of the appropriate metallicity. The parameter space of the survey stars that extend beyond that of the training data  extrapolates directly along the isochrones to the lowest \logg\ and \teff\ values.  
The stellar parameters returned by \tc\ also compare well with the results from \aspcap, where the ranges overlap. For the $\approx$ 18,000 stars in the 
range $(-5^\circ < l < 65^\circ, |b| < 15^\circ)$ for which \apogee\ stellar parameters are available from \aspcap,  \tc's parameters are identical within:  $\Delta$\feh\ = --0.04 $\pm$ 0.07, $\Delta$\alphafe\ = 0.01 $\pm$ 0.06 , $\Delta$\teff\ = -22 $\pm$ 55~K, and $\Delta$\logg = 0.02 $\pm$ 0.12. }

The stellar parameters for these stars, their $\chi^2_{reduced}$ values and their corresponding distances as well as velocities are provided in Table 1 in the Appendix (and in full online). Stars with a poor fit to the model $\chi^2_{reduced}$ $>$ 6, were excluded from analysis (approximately 5\% of the stars). 

We do include the stars in this region of extrapolated stellar parameter space in our analysis of the kinematics as stars as a function of metallicity. However, we use only coarse bins in metallicity, of 0.5 dex intervals (see in Section 5). This allows us to compare the kinematic behavior of metal-rich and metal-poor stars. The alternatives, of either eliminating all stars in the region of extrapolated label-space, or using only the stars with stellar parameters provided by \apogee\ (which are also in extrapolated label-space for \aspcap), produces qualitatively the same results as those obtained with their inclusion. However, we have the benefit of working with a larger number of stars using the full range of \tc's results. 

To isolate the stars in the region of the bulge, we use the stellar parameters and obtain distances by shifting the stars in \teff\ and \logg\ space to their nearest position on a 10~Gyr isochrone \citep{Bressan2012}, for an isochrone at the metallicity of the star by interpolating between isochrones. The apparent {dereddened} magnitude of each star was calculated by adopting the RJCE-WISE extinction value for that line of sight, provided in the \apogee\ DR12 data \citep{Majewski2011,Zasowski2013}. In the few cases, for 160 stars, where no reddening estimates were available, a value of $A(K_s) = 0$ was adopted, which effectively provides an upper limit to the distance estimate from the isochrone matching. The adopted distance error on each star is 30\%. These errors are composed of uncertainties in the reddening, which is most significant in the plane, uncertainties on the stellar parameters (which are $\Delta$(\teff, \logg, \feh, \alphafe) = 70~K, 0.2, 0.1, 0.05 dex)), and also any systematic offsets in the stellar parameters or else the isochrones. 

\begin{figure*}
\centering
        \includegraphics[scale=0.45]{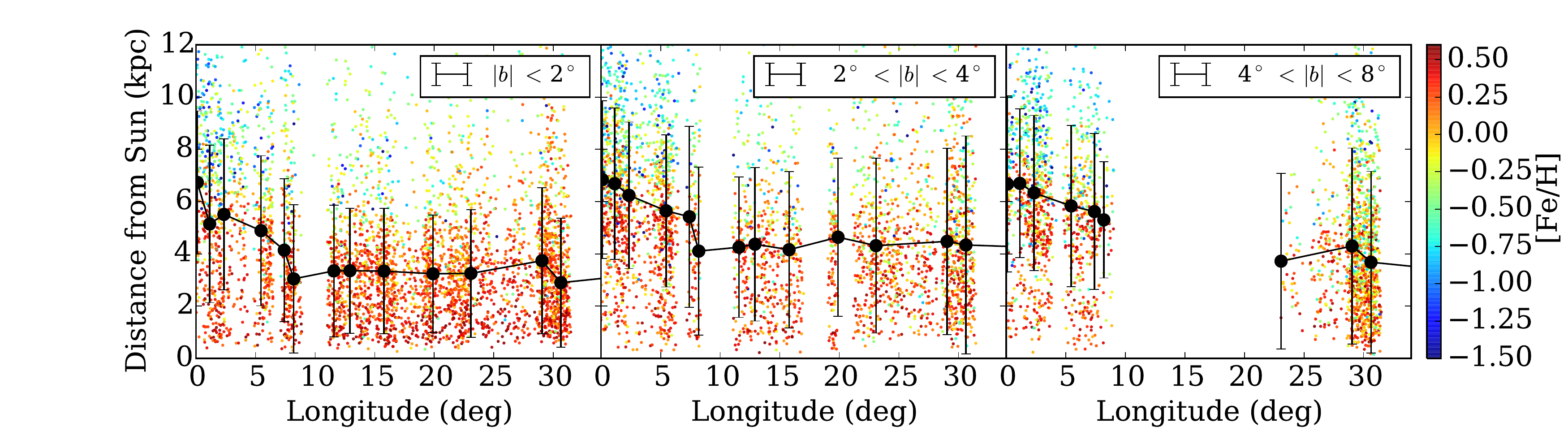}
\caption{{The median distance showing the $1-\sigma$ dispersion in the errorbars, of the \apogee\ stars as a function of longitude, for $|b|$ $<$ 2$^\circ$ at left, 2$^\circ$ $<$ $|b|$ $<$ 4$^\circ$, centre  and 4$^\circ$ $<$ $|b|$ $<$ 8$^\circ$, at right.}}
\label{fig:prior}
\end{figure*}

{Figure \ref{fig:prior} shows the distance distribution of stars across longitude, colored by metallicity, for three slices in latitude. This shows that using distances calculated directly from the isochrones, the stars in the \apogee\ sample are biased to be located on the near side of the bulge and there is a strong bias in metallicity, with few metal-rich stars beyond \rgc\ $>$  8kpc. In addition to the large uncertainties in distance for the red giant selection ($\approx$ 30\%), particularly in highly extincted regions,  there may be systematic offsets in the distance and the median distance may be underestimated by this simple isochrone distance calculation. We examine the chemo-kinematics of stars well in front of the bulge ($<$ 3 kpc) and for our bulge selection  (4 -- 12 kpc) in Sections 3-5 to validate that the relative scale of our distances effectively enables the foreground stars, which show disk-like kinematics and chemistry that is dissimilar to the bulge sample, to be effectively eliminated from our analysis. Systematic distance offsets may be a consequence of the stellar parameter scale (e.g. temperature scale) of the stars or else overestimated values of extinction, particularly in the highly reddened plane region ($|b|$ $<$ 4$^\circ$). The metallicy bias as a function of distance in the \apogee\ sample is discussed in Garcia-Perez et al., 2016 (in prep). }

{The bias in distance as a function of \feh\ and also the concentration in distance to the near side of the bulge  is present even where the extinction is smaller (above $|b|$ $>$ 4$^\circ$). The peak of the overall distance distribution is similar to that found by \citet{Ness2013a} for red clump stars, at around 5--6 kpc from the Sun although the \apogee\ stars and particularly the metal-rich stars do not reach far beyond 8 kpc. }

{For our distance cuts in our selection of bulge stars, we adopt a broad heliocentric distance criterion of $4 - 12$~kpc for bulge membership, which includes stars from the end of the bar at $l$ = 30$^\circ$ to the far side of the bar at $l$ $<$ 0$^\circ$.  {Although broad, this distance selection effectively and critically eliminates many of the stars in the foreground disk along the line of sight to the bulge whilst including the stars in the bulge, the vast majority of which are on the near side}.  Our distance estimates are included with the stellar parameters for DR12 in Table \ref{tab:online} in the Appendix. }

Our distance cut of $4 - 12$~kpc provides sufficient margin to  account for uncertainties in the distance determination, whilst minimizing the contamination of the foreground objects. Many previous bulge studies \citep[e.g.,][]{zoccali2008, howard2008} {do not determine distances to eliminate foreground objects; however these studies do select stars in a confined region of the color magnitude diagram to maximise the likely true bulge stars in the sample}. We find with our determined distances, and \apogee\ colour cuts \citep{Zasowski2013}, that approximately 35\% of the sample are foreground objects along the line of sight toward the bulge located at distances $<$ 4 kpc (which we correspondingly eliminate) \citep[a similar fraction as found by][]{Ness2013a}. The distances in the plane show bias to distances nearer to the Sun, and the median distances increase above the plane ($|b|$ $>$ 2$^\circ$) in regions of lower extinction, where the fainter stars are located at larger distances.

\begin{figure*}
    \begin{subfigure}[b]{0.5\textwidth}
    \centering
    \includegraphics[width=1.15\linewidth]{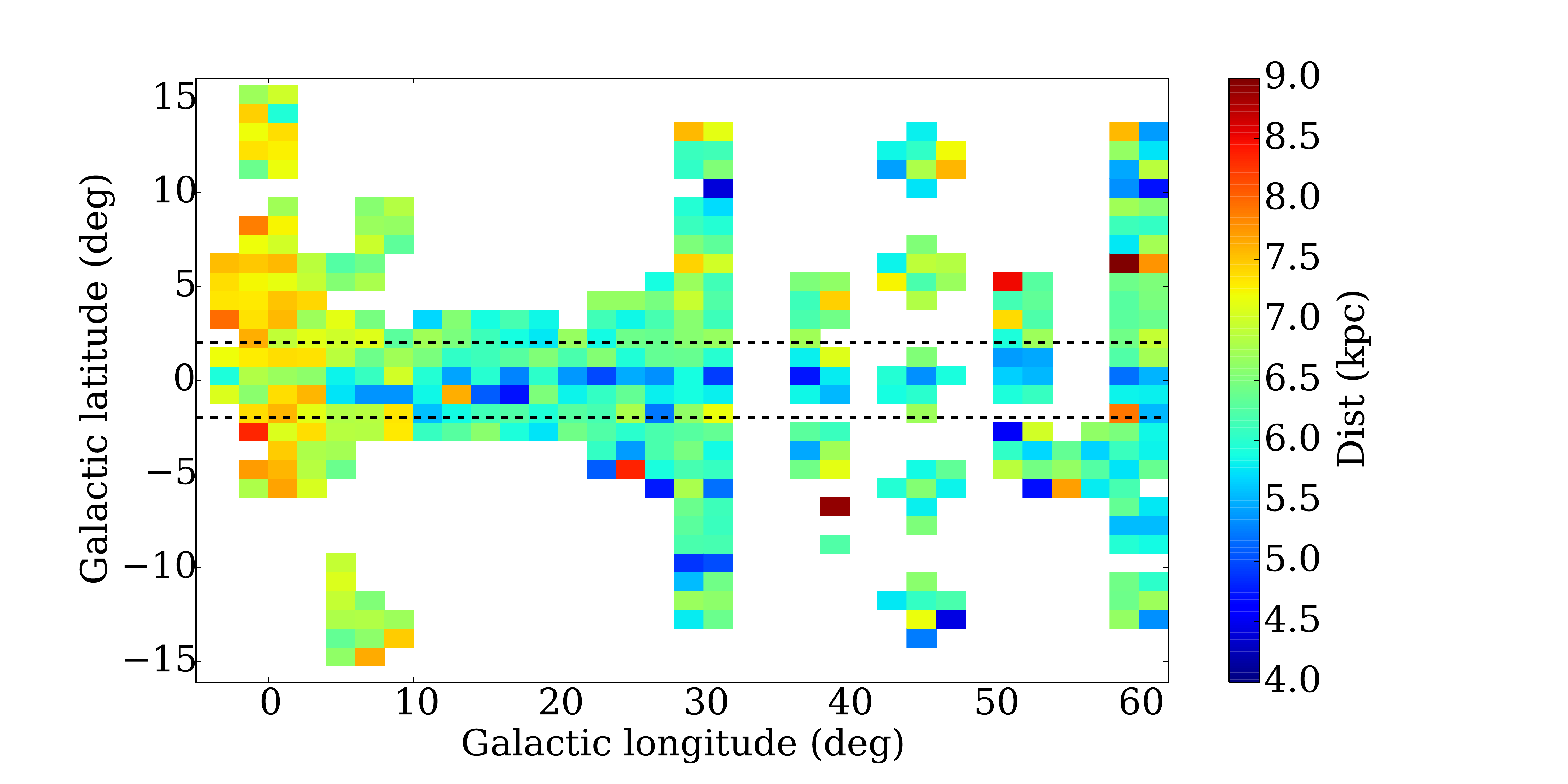} 
\caption{ }
\label{fig:distnuma}
  \end{subfigure}%
  \begin{subfigure}[b]{0.5\textwidth}
    \centering
    \includegraphics[width=1.15\linewidth]{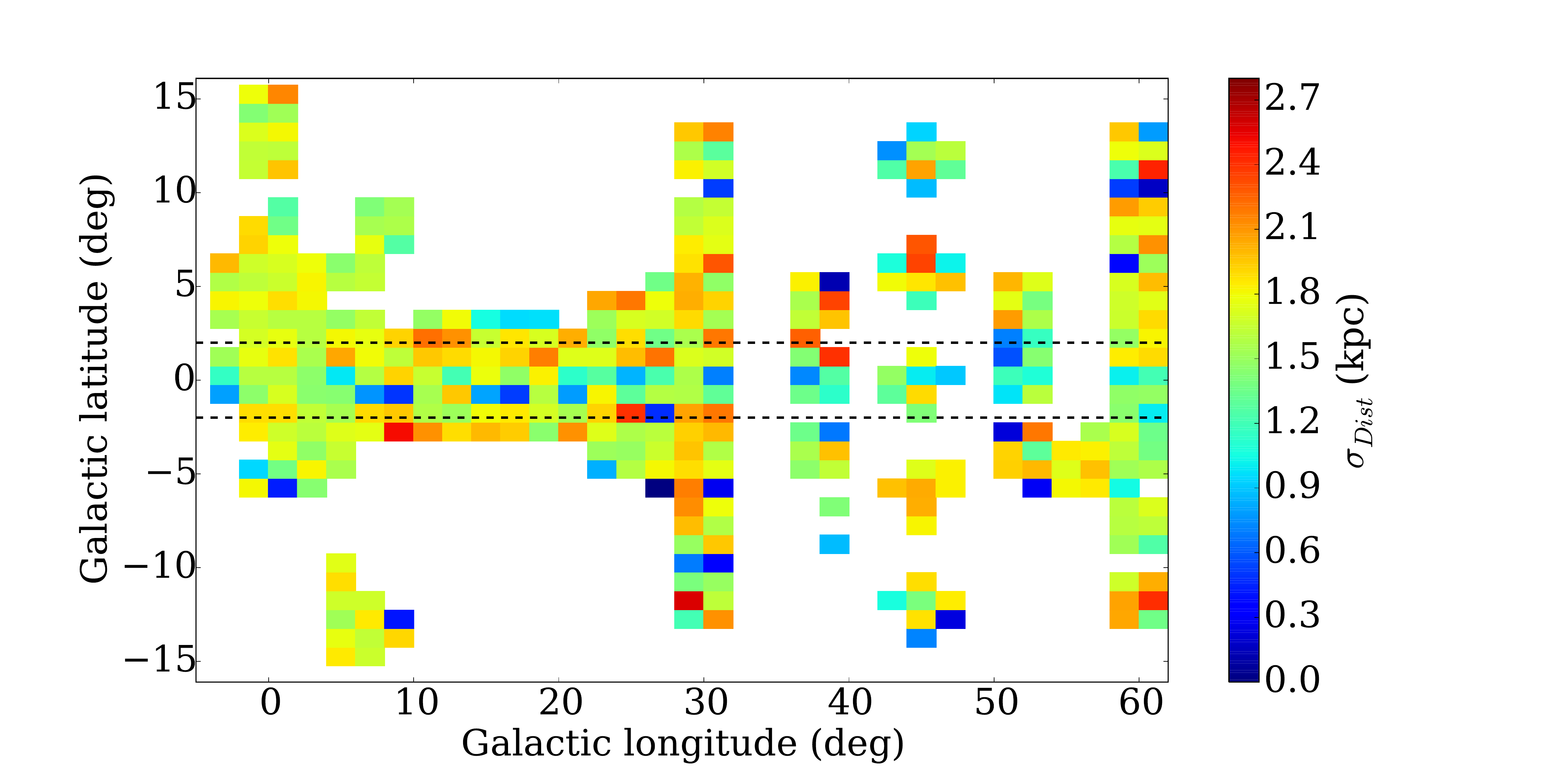}
    \caption{ }
\label{fig:distnumb}
  \end{subfigure}
   \caption{{(a) Mean distance and (b) distance dispersion maps for the 10,000 \apogee\ bulge and disk stars, isolated with a heliocentric distance cut of 4 -- 12kpc. Note that the stars in the plane are nearer to the Sun, and that the range of distances spanned at higher latitudes is largest above the plane }}
   \label{fig:dist}
\end{figure*}

{Figure \ref{fig:distnuma} shows the mean distance distribution across $(l,b)$ with our distance cuts implemented; off the plane ($|b|$ $>$ 2$^\circ$), the mean distances are around 6.5--7 kpc from the Sun and Figure \ref{fig:dist}b shows that the distance distribution is broader at higher latitudes.    Adopting instead the distance estimates of \citet{Hayden2015} for our stars and repeating the analyses below, we obtain no significant differences in our results and we draw no different conclusions to those in this analysis with \tc-derived parameters and our own distances. }

\begin{figure}[h!]
\centering
    \includegraphics[scale=0.2]{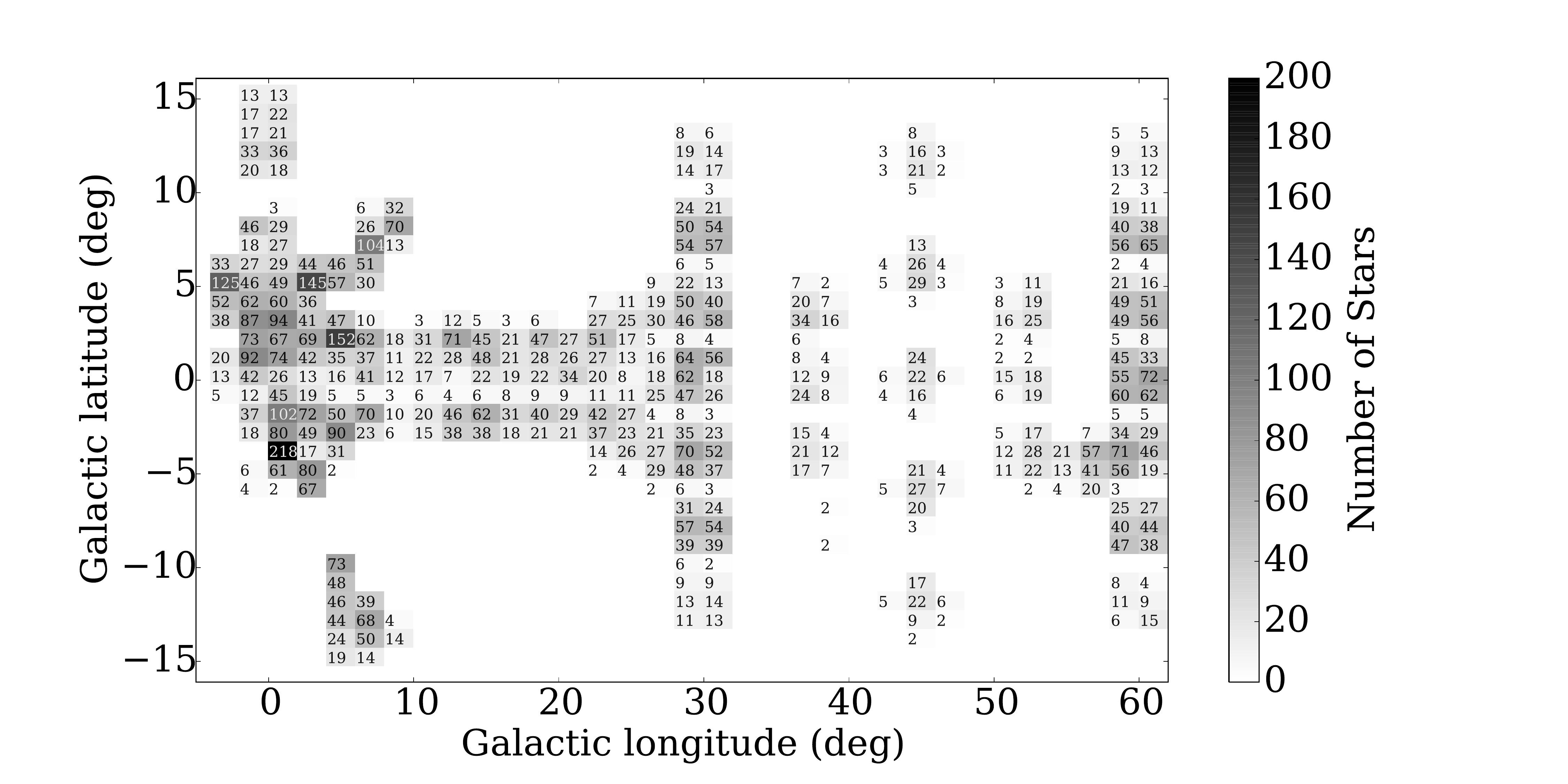}
\caption{The numbers of stars in each bin used for the determination of the mean velocity and velocity dispersion maps, for stars with distances = 4 -- 12 kpc. There are a total of $\approx$ 10,000 stars at these distances for these fields.  }
\label{fig:stars2}
\end{figure}

After our distance cuts, our sample of bulge and inner disk stars is reduced to $\sim$ 10,000. For these stars, we examine the mean velocity and dispersion by mapping across bins with $\Delta l = 2^\circ$ and $\Delta b = 1^\circ$ across the bulge and inner disk (Section~\ref{sec:kinematic_maps_all}). The typical number of stars in each of these bins is $>$ 20 as shown in Figure \ref{fig:stars2}.

%\begin{figure*}
%%\centering
%\flushright
%    \includegraphics[scale=0.33]{numbers_noprior2.pdf}
%\caption{The numbers of stars in each bin used for the determination of the mean velocity and velocity dispersion maps, for stars with distances = 4 -- 12 kpc. There are a total of $\approx$ 10,000 stars at these distances for these fields.  }
%\label{fig:stars2}
%\end{figure*}

%
\section{Kinematic Maps of the Milky Way} \label{sec:kinematic_maps_all}

\begin{figure*}
    \begin{subfigure}[b]{1.1\textwidth}
    \centering
    \includegraphics[width=0.8\linewidth]{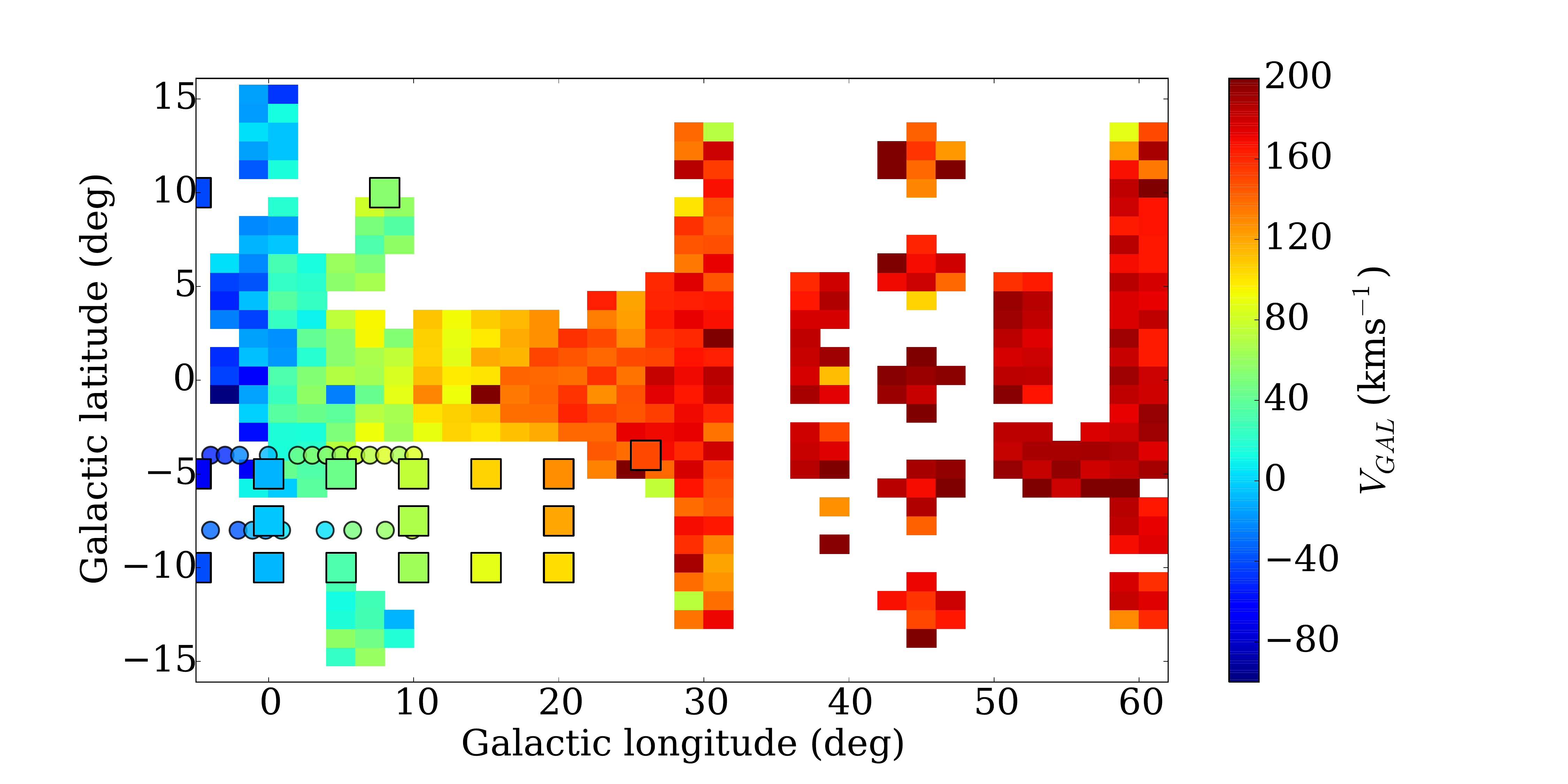}
\caption{}
\label{fig:rotstda}
  \end{subfigure}%
  
  \begin{subfigure}[b]{1.1\textwidth}
    \centering
    \includegraphics[width=0.8\linewidth]{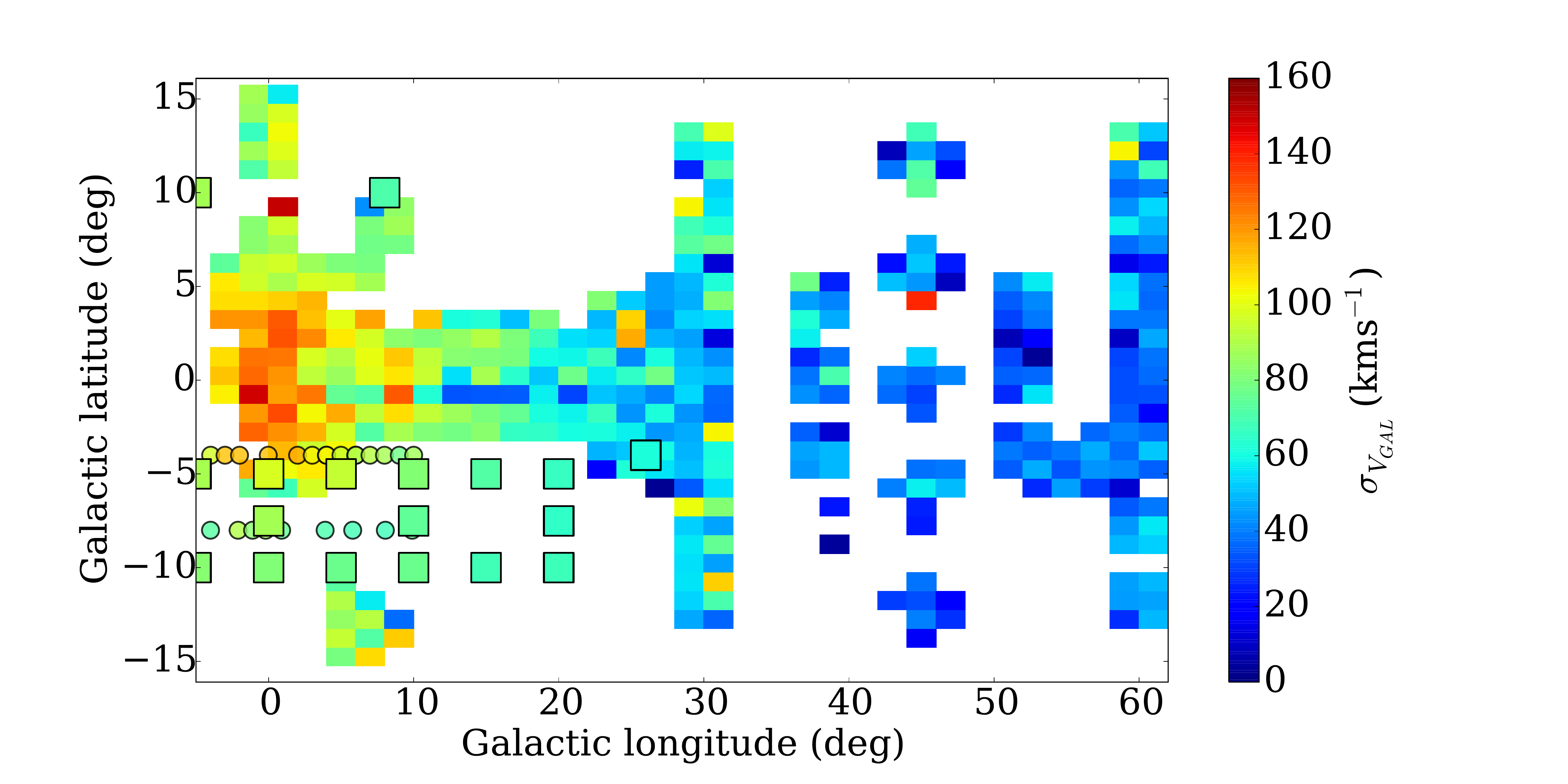}
\caption{} 
\label{fig:rotstdb}
  \end{subfigure}
  \caption{Rotation (a) and velocity dispersion (b) maps for the 10,000 \apogee\ bulge and disk stars, taken within a distance cut of 4 -- 12kpc. The squares with black outlines are from the ARGOS data, and contain about 600 stars. The circles with black outlines are BRAVA fields and contain between 100-200 stars in each of these fields. The number of stars in each of the APOGEE fields is given in Figure \ref{fig:stars2}. Typical errors in the velocities for individual \apogee\ stars are $<$ 1 kms$^{-1}$. These maps show a cylindrical rotation that is symmetrical about the major axis and a dispersion profile that is hottest at the centre, similarly to the kinematic profiles of boxy bulges other galaxies. There is also good agreement {between \apogee, ARGOS and BRAVA results.}  }
  \label{fig:rotstd}
\end{figure*}

The rotation and dispersion maps of the bulge and inner disk, comprising the 10,000 stars spanning estimated distances of 4--12 kpc, are presented in Figure \ref{fig:rotstd}. The colormap shows the mean velocity and the dispersion in each bin. Figure \ref{fig:rotstd}a demonstrates that the rotation curve appears cylindrical into the plane, with about the same average velocity across slices in latitude. Figure \ref{fig:rotstd}b reveals that bulge is kinematically hottest at the very centre and the dispersion decreases with $(l,b)$ and flattens as a function of $b$ at the disk $l$ $>$ 15$^\circ$.  The comparable results from BRAVA \citep[][]{howard2008,kunder2012} and ARGOS \citep[][]{Freeman2012,Ness2013b} are also included on these maps. The \apogee\ velocity data agree well with the results from the BRAVA and ARGOS surveys, which show comparable mean rotation and dispersion values for fields that overlap and a near cylindrical rotation profile across high to low latitudes. The rotation and dispersion profiles transition smoothly into the disk and do not reveal any large-scale dynamical substructure present in the bulge. 

\begin{figure}[h]
\centering
    \includegraphics[scale=0.2]{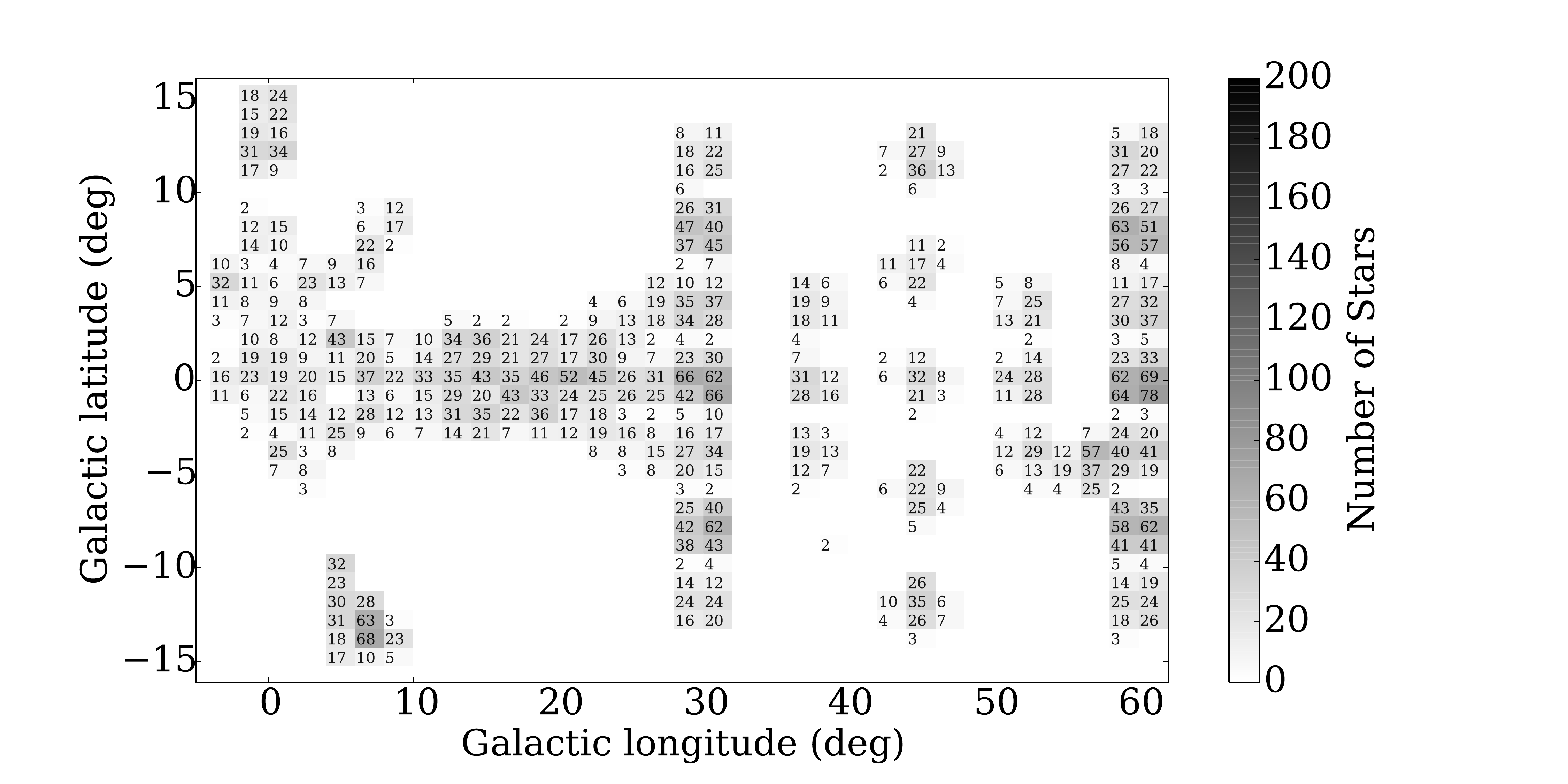}
\caption{The numbers of stars in each bin used for the determination of the mean velocity and velocity dispersion maps, for stars with distances = 0 -- 3 kpc. There are a total of $\approx$ 6800 stars at these distances for these fields.  }
\label{fig:near_nums}
\end{figure}
%\begin{figure*}
%%\centering
%\flushright
%    \includegraphics[scale=0.33]{numbers_near_noprior2.pdf}
%\caption{The numbers of stars in each bin used for the determination of the mean velocity and velocity dispersion maps, for stars with distances = 0 -- 3 kpc. There are a total of $\approx$ 6800 stars at these distances for these fields.  }
%\label{fig:near_nums}
%\end{figure*}

Overall, these maps are remarkably similar to the cylindrical rotation and velocity dispersion profile of barred galaxies \citep[e.g., NGC 7332 by][]{FalconB2004} and also the corresponding maps of N-body simulations which evolve boxy bulges from the disk \citep[e.g.][]{Athanassoula2002}. Two critical differences between the IFU maps of external galaxies and our maps is that we have eliminated the foreground objects, and that these averages are from a star-by-star analysis, rather than an integrated stellar population. 

\begin{figure*}
    \begin{subfigure}[b]{0.5\textwidth}
    \centering
    \includegraphics[width=1.1\linewidth]{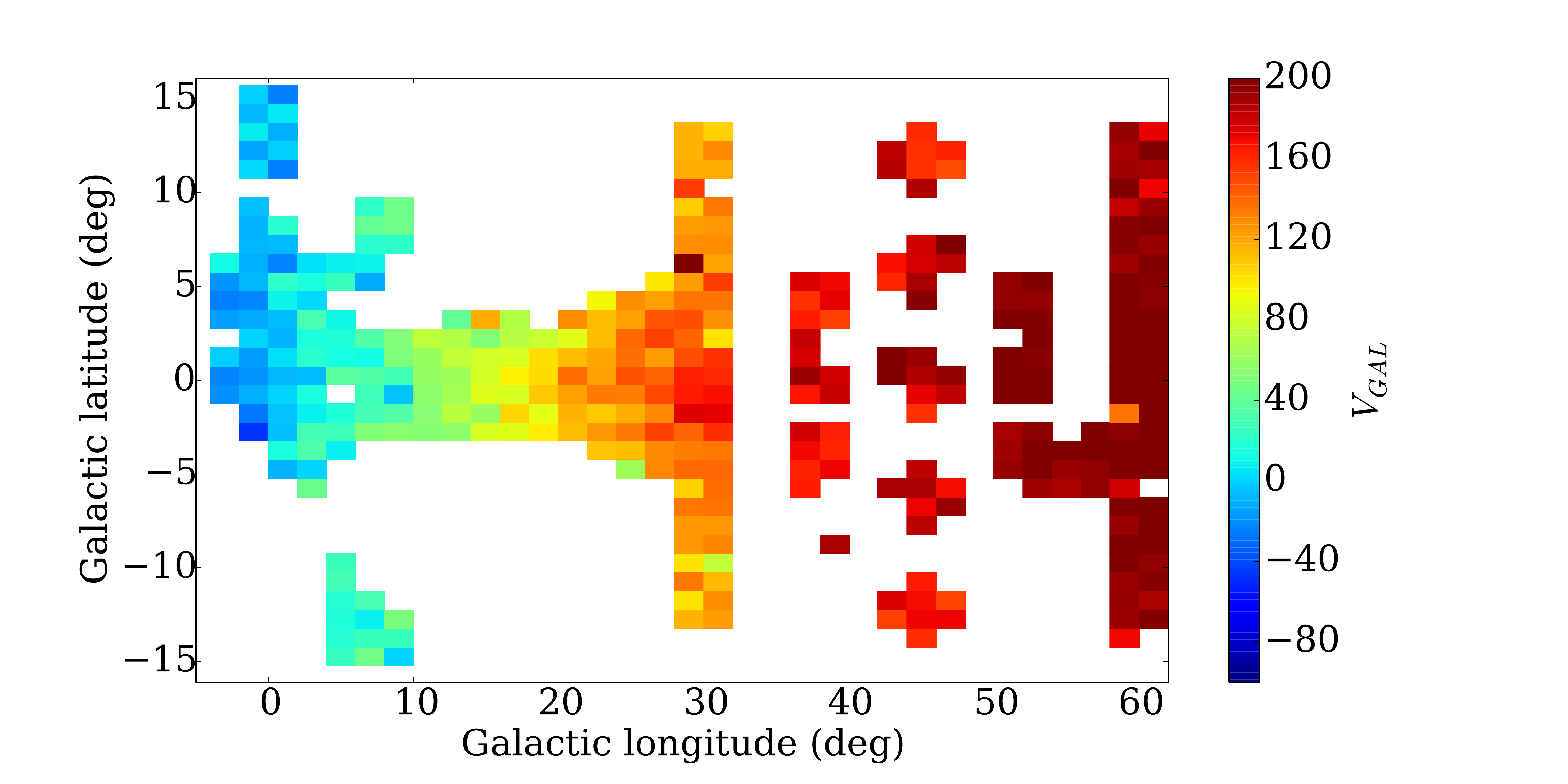} 
\caption{}
  \end{subfigure}%
 \begin{subfigure}[b]{0.5\textwidth}
    \centering
    \includegraphics[width=1.1\linewidth]{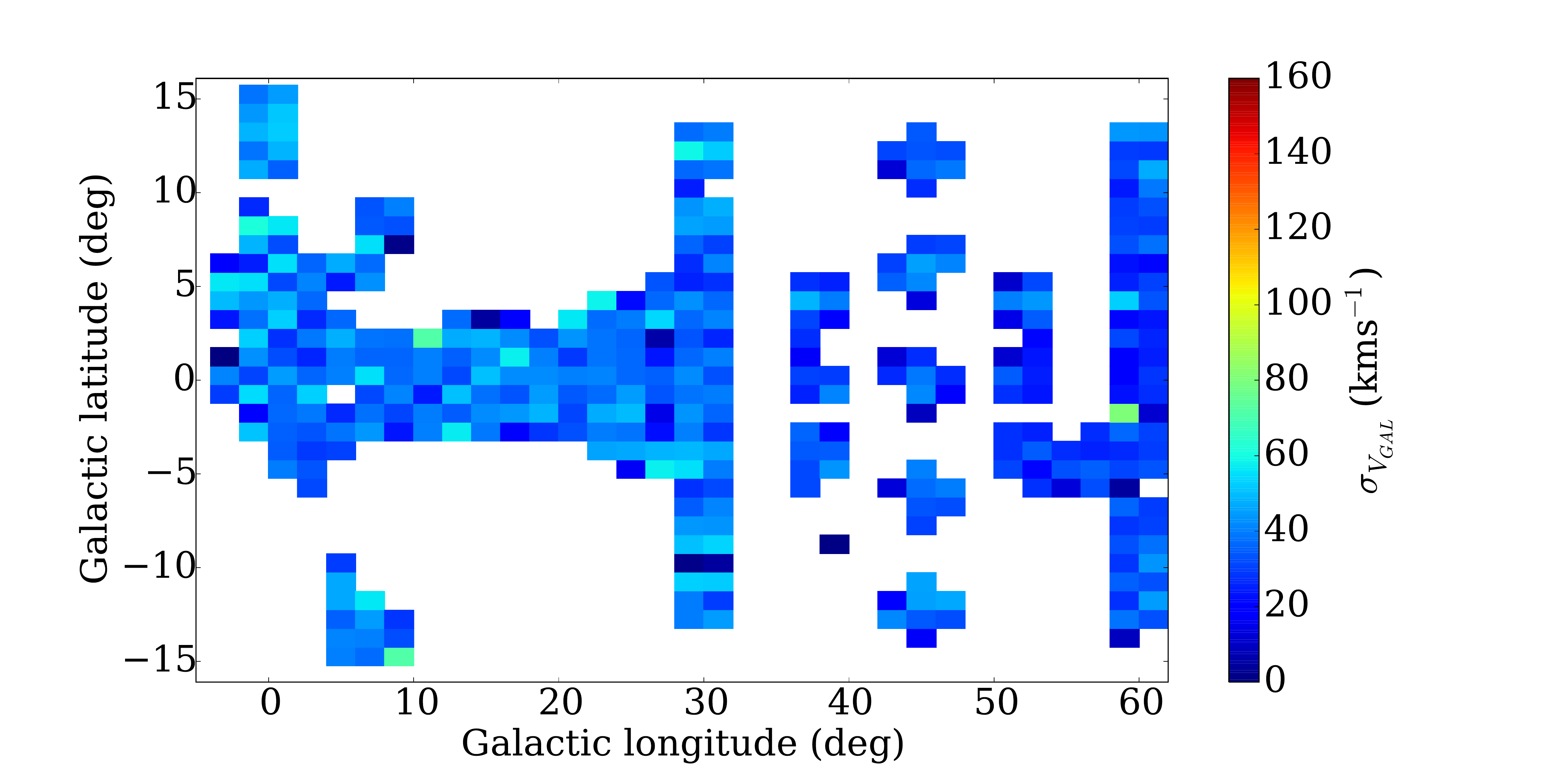} 
\caption{} 
\end{subfigure}
\caption{Similar to Figure~\ref{fig:rotstd} but for stars at distances from the Sun smaller than 3 kpc: (a) Rotation and (b) Velocity dispersion maps for the 6800 presumed disk stars in the foreground of the bulge, with distances $<$ 3 kpc. }
  \label{fig:near}
\end{figure*}

We examine the comparable kinematic profile for the 6800 foreground objects, which are shown in number distribution in Figure~\ref{fig:near_nums}, where stars have been selected with distances from the Sun $<$ 3 kpc.  The rotation of the disk stars nearer the Sun presented in Figure \ref{fig:near}a is similar to the stars at the distance of the bulge, although the rotation is slower. However, the velocity dispersion shown in Figure \ref{fig:near}b is clearly dissimilar to that of the bulge population, with a flatter, low-dispersion profile from inner to outer longitudes, as expected for a foreground disk population. 

\section{Comparison to an N-body model} 

\subsection{Rotation and velocity dispersion profiles of the bulge} 

The overall kinematic trends from the ARGOS survey, which comprised 28 fields of about 600 stars at distances of the bulge (4.5 -- 11.5 kpc from the Sun), across $b$ = --5$^\circ$, --7.5$^\circ$ and --10$^\circ$ and longitudes $|l|$ $<$ 31$^\circ$,  are a good match to N-body models of boxy/peanut bulges for stars with metallicities of \feh\ $>$ --0.5 \citep{Ness2013b}.  In this section, we compare the global APOGEE kinematic patterns to a model of Athanassoula (2005,2007), which has been scaled to match the ARGOS data in \citet{Ness2013b}.

This model consists of a disc and dark mater halo with density distributions as described in \citet{Athanassoula2007}. The model is evolved from a thin disk that forms a bar at early times, whose instabilities later form a boxy/peanut bulge.  Because the model contains no gas, there is no star formation and there is no information on chemistry of the stars. The model has a Toomre parameter of $Q = 1.2$ and a vertical height that is 0.2 times the disc scale length. This model is scaled to the spatial size of the COBE bar. \citet{Ness2013b} verified this model to be a good match to the ARGOS kinematics at latitudes {below} the plane, from $b$ = --5$^\circ$ to --10$^\circ$. 

This model was compared in \citet{Ness2013b} to the ARGOS data across latitudes of $b$ = $-5, -7.5$ and $-10^\circ$ and found to be a good qualitative match to the data. The \apogee\ data has stars across a broader spatial region of the bulge, and critically, in the plane. Figure \ref{fig:ap_ar} shows the comparison of the model with the  ARGOS fields across the three latitudes \textit{plus} the \apogee\ data in the plane, where \apogee\ stars that are within $b$ $<$ 2$^\circ$ have been binned to comprise the in-plane measurements. {The bar in the model has been rotated to 27$^\circ$ and stars in the model have been selected at distances 4 -- 12 kpc from the Sun, similar to the \apogee\ and ARGOS data}.

 There are small quantitative differences {between data and model seen in Figure~\ref{fig:ap_ar}}, but the trends are matched very well by the model. Particularly, that inside the boxy bulge region, the dispersion is peaked at low latitudes and latitude dependent. At $b$ = 0$^\circ$, the dispersion in the plane is almost double the dispersion at b=10$^\circ$. Outside of $l$ $>$ 10$^\circ$, however, the disk shows comparable dispersion across latitude, at b=0$^\circ$ and b =10$^\circ$.
 
 \subsection{The velocity and velocity dispersion maps of the bulge} 
 
 \begin{figure*}
\flushright
    \includegraphics[scale=0.43]{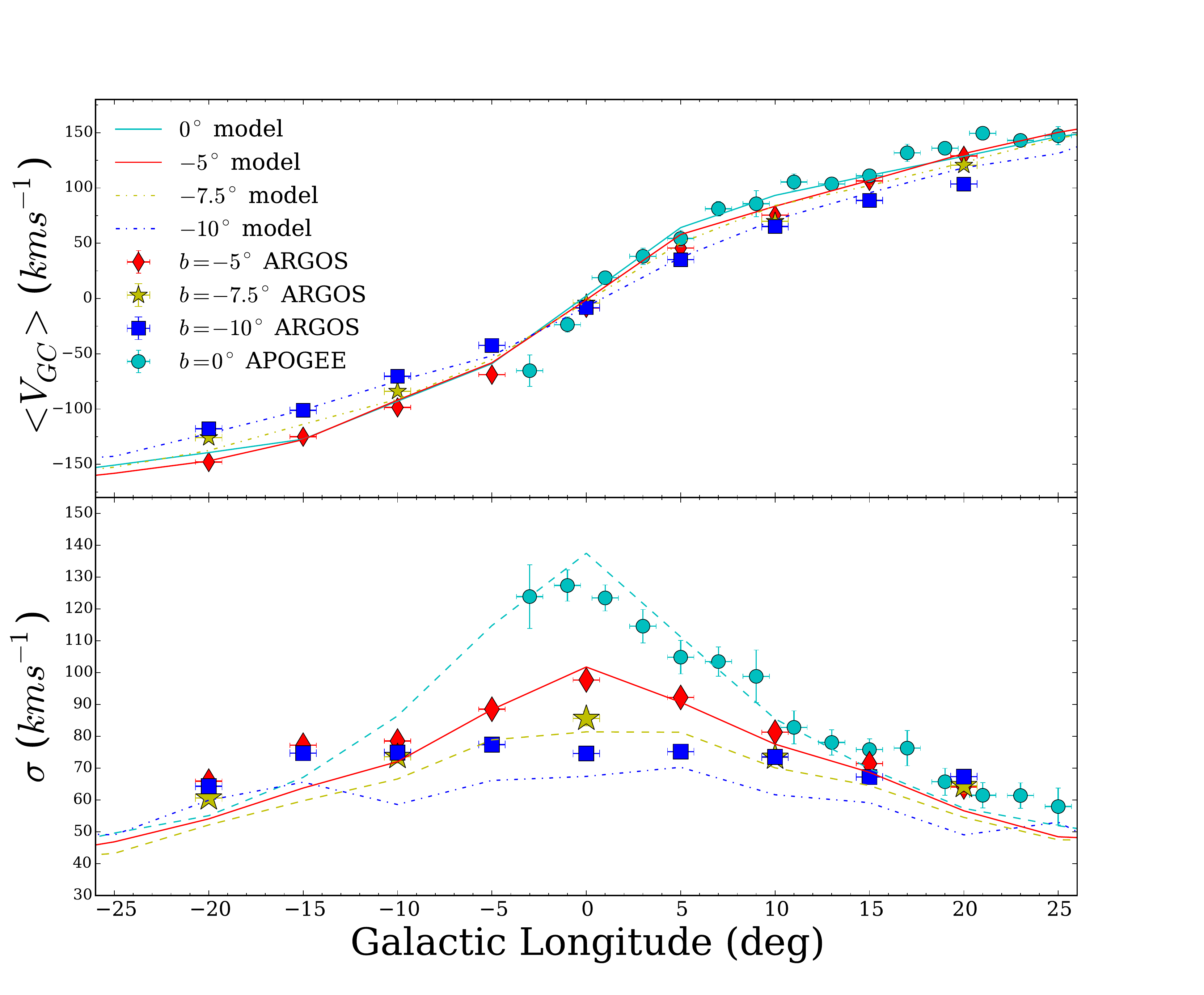}
\caption{The velocity and velocity dispersion for the bulge stars from the ARGOS survey for fields at $b$ = --5, --7.5$^\circ$ and --10$^\circ$ (red, yellow and blue symbols) and the \apogee\ survey, for stars in the plane within $|b|$ $<$ 2$^\circ$ (cyan symbols). Stars are located at distances of 4 -- 12 kpc. The model used here and in \citet{Ness2013b} is shown by the lines. The model is rotated so that the bar is 27$^\circ$ with respect to the line of sight and provides a good qualitative match in velocity rotation and dispersion for both ARGOS and \apogee\ data, with a triangularly peaked dispersion at low latitudes and a flatter dispersion at high latitudes, and a comparable, latitude independent dispersion outside of the bulge in the disk. The rotation is only weakly dependent on latitude: the rotation at $b=0^\circ$ is only marginally faster than at $b = -10^\circ$, for both data and model.}
\label{fig:ap_ar}
\end{figure*}

\begin{figure*}
  \begin{subfigure}[b]{0.5\textwidth}
  \flushright
    \includegraphics[width=1.1\linewidth]{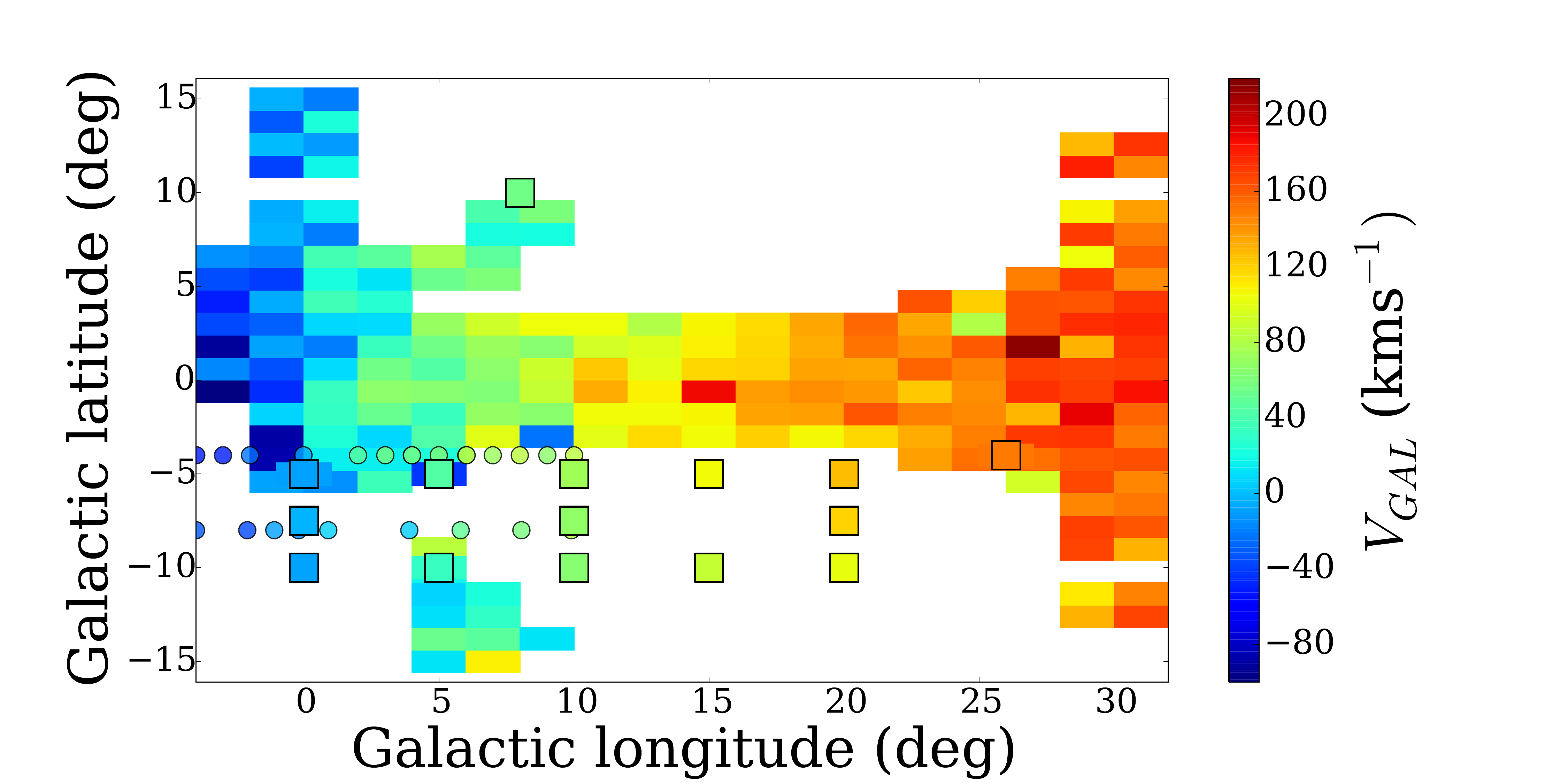}
    \caption{}
  \end{subfigure}%
     \begin{subfigure}[b]{0.5\textwidth}
       \flushright
    \includegraphics[width=1.1\linewidth]{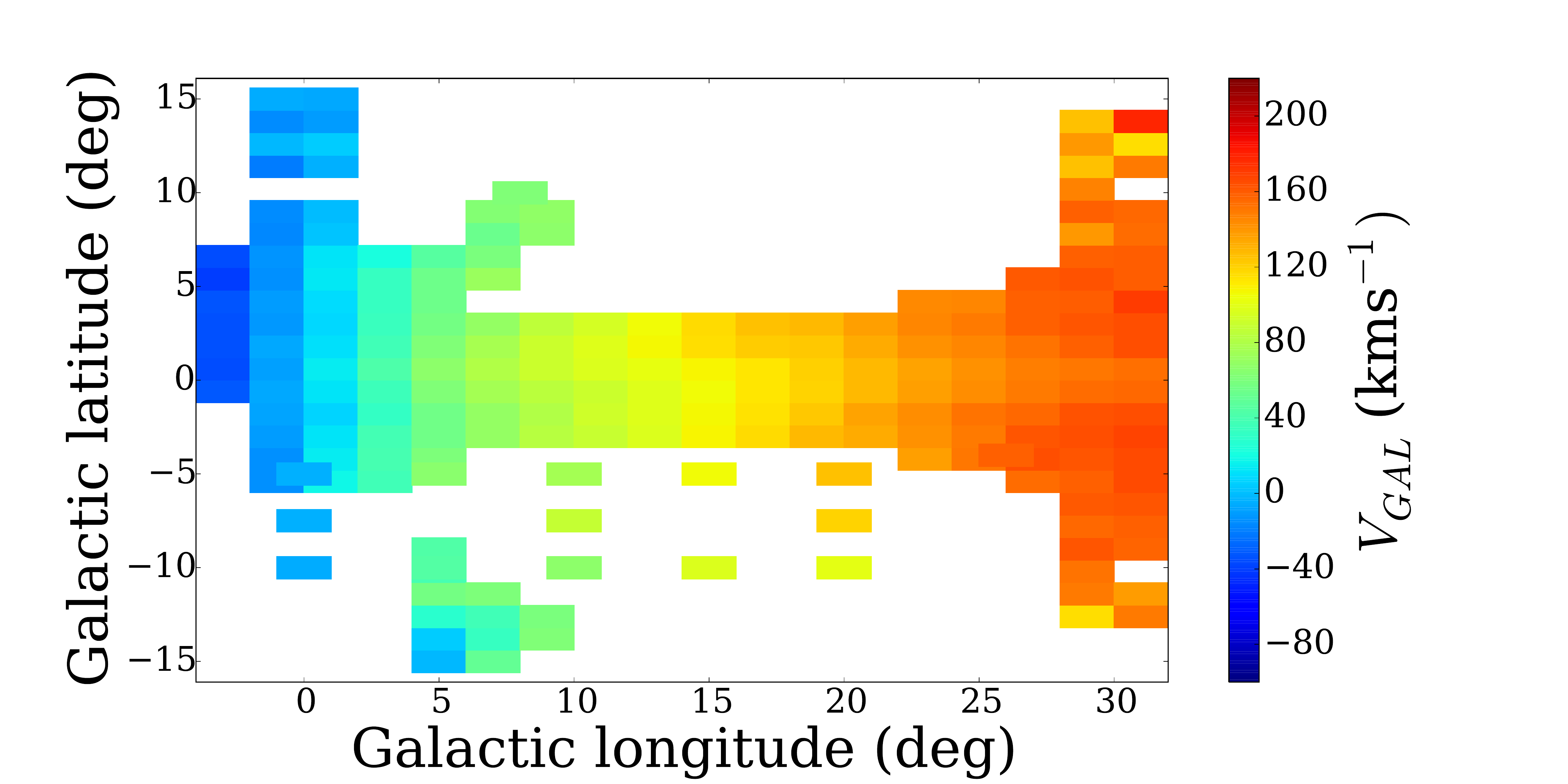} 
    \caption{}
    \end{subfigure}
   \quad
   \begin{subfigure}[b]{0.5\linewidth}
     \flushright
   \includegraphics[width=1.1\linewidth]{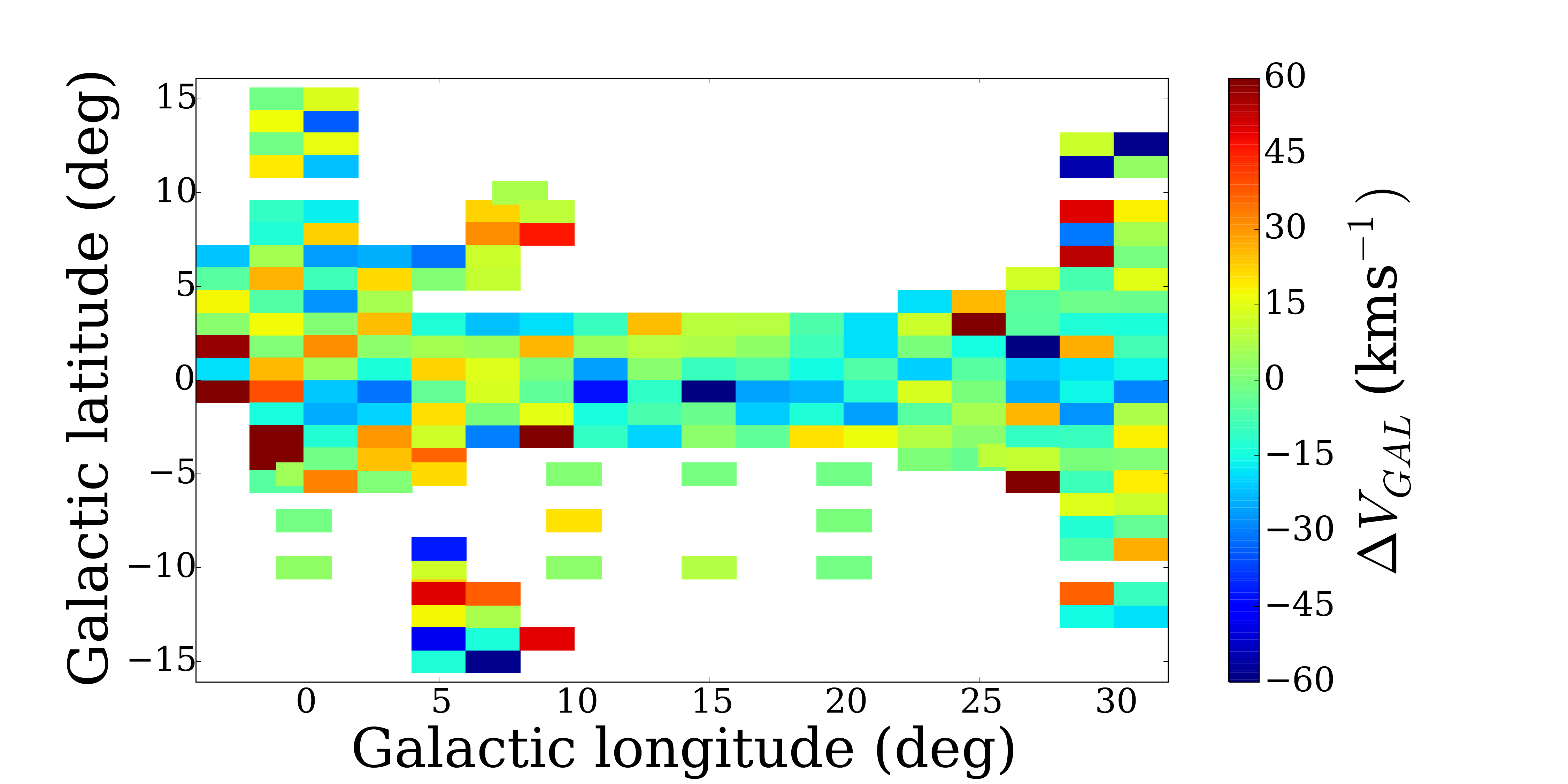}  
    \caption{}
  \end{subfigure}
    \begin{subfigure}[b]{0.5\linewidth}
      \flushright
         \includegraphics[width=1.1\linewidth]{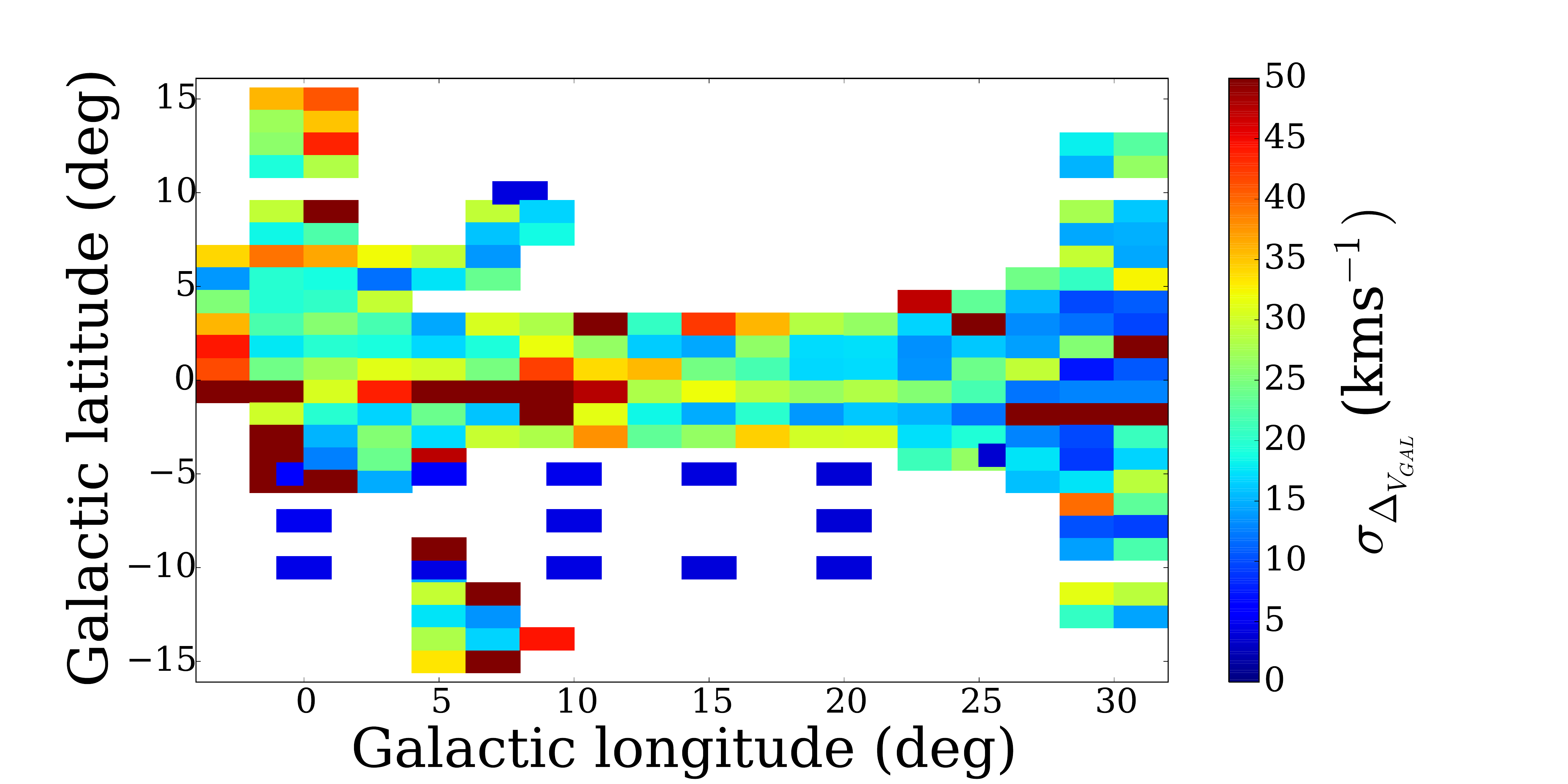} 
    \caption{}
  \end{subfigure}%
  \caption{Comparison of mean velocity maps from data and a barred galaxy model. 
  (a) Compilation of mean velocity measurements from the APOGEE, ARGOS, and BRAVA surveys: distance cuts are implemented for ARGOS and \apogee\ data, taking stars 4 -- 12 kpc from the Sun. ARGOS data are in the larger square outlined bins and BRAVA data are in the circular fields. 
  (b) Mean velocity measurements in the barred galaxy model from \citet{Ness2013b} described in \citet{Athanassoula2003, Athanassoula2007} for stars 4 -- 12 kpc from the Sun.
  (c) {Differences between the model in panel \textit{(b)} and the data in panel \textit{(a)}; these do not show coherent structure, indicating overall the model well matches the data}.
  (d) Observational errors on the differences in panel \textit{{(c)}} to indicate where there are few stars per bin and where large differences in \textit{(c)} have corresponding large errors.}
 \label{fig:model1}
\end{figure*}

\begin{figure*}
  \begin{subfigure}[b]{0.5\linewidth}
    \flushright
    \includegraphics[width=1.1\linewidth]{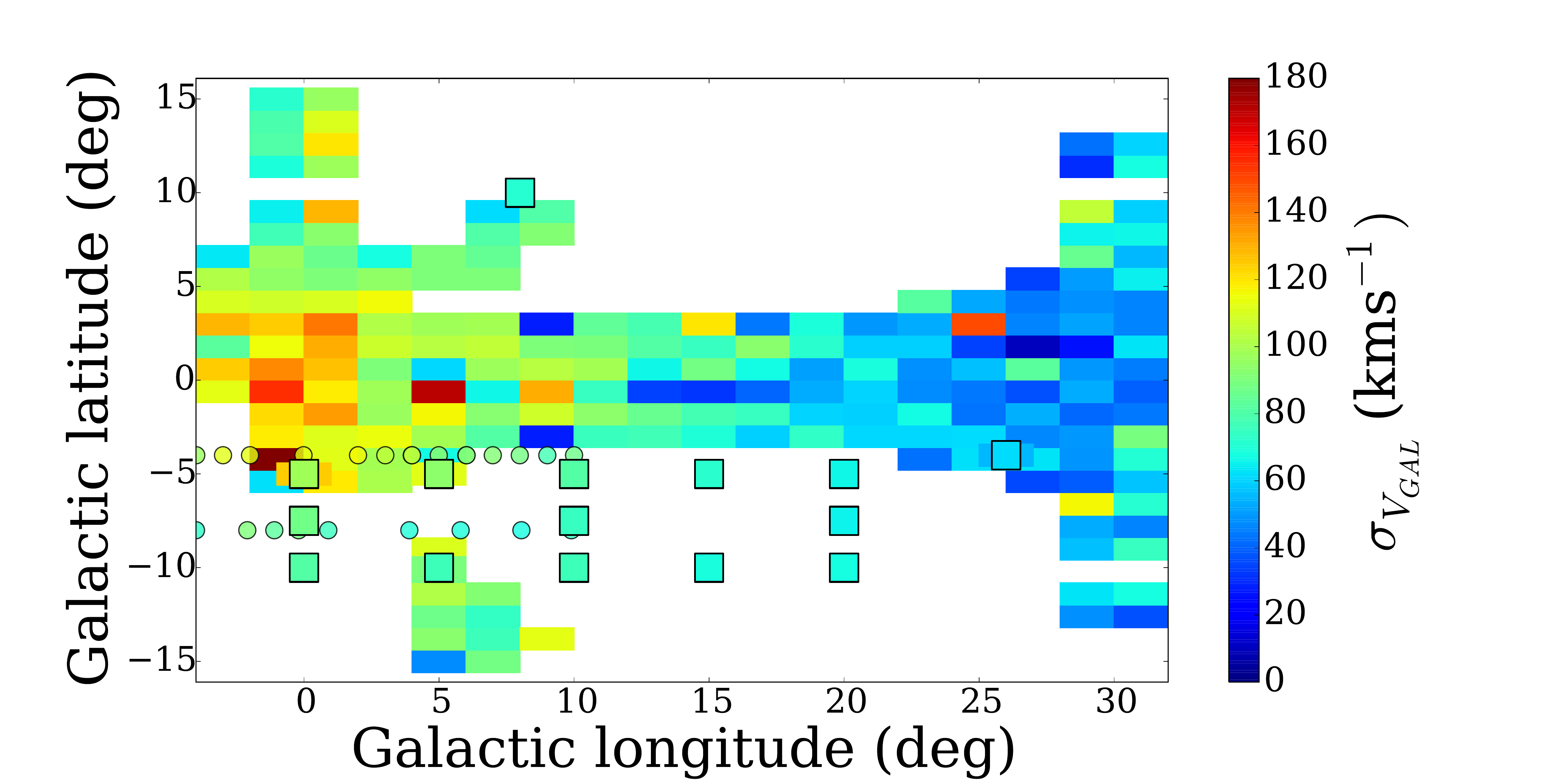}
    \caption{}
  \end{subfigure}%
  \begin{subfigure}[b]{0.5\linewidth}
    \flushright
    \includegraphics[width=1.1\linewidth]{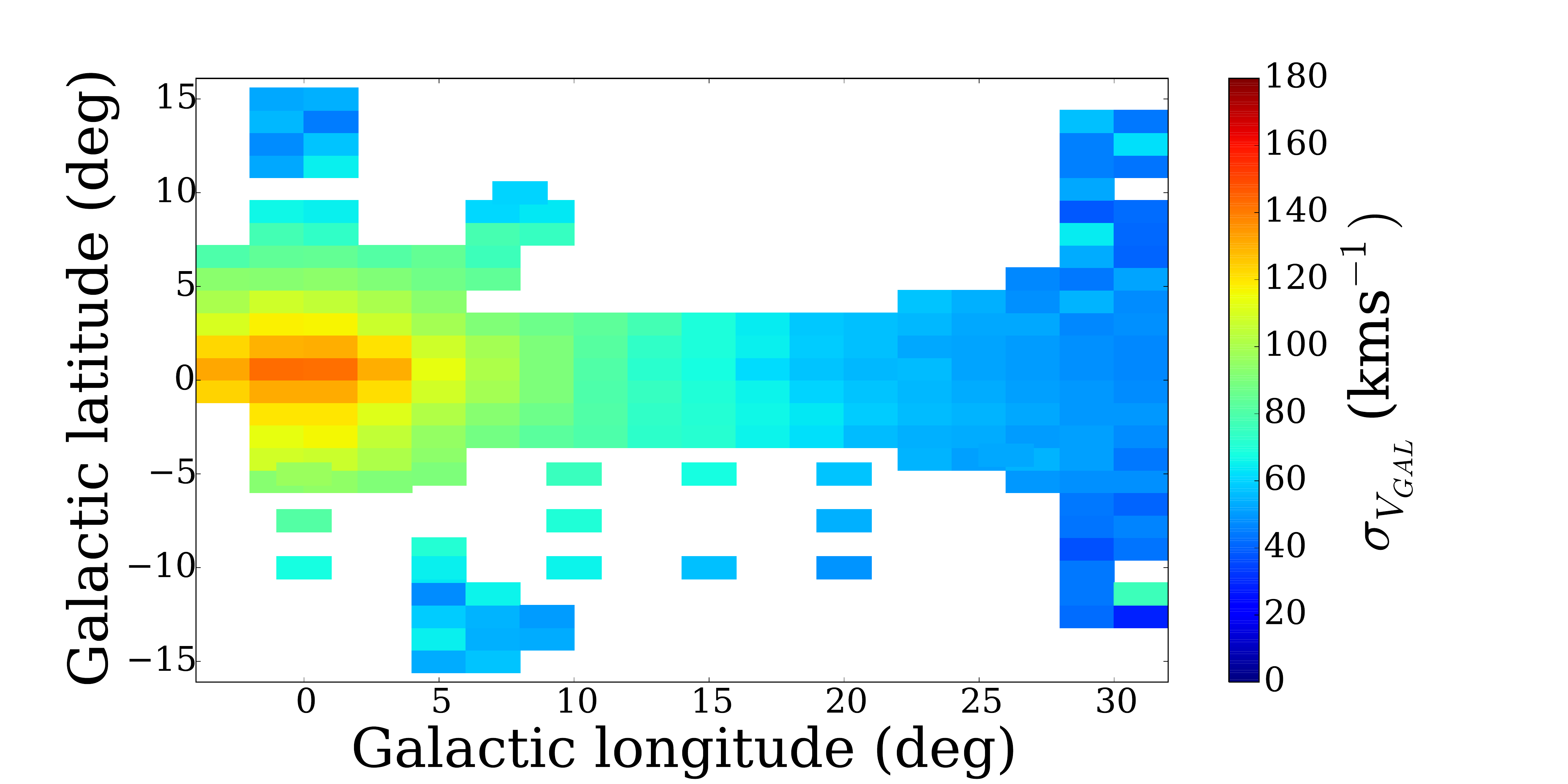}
    \caption{}
  \end{subfigure}
      \begin{subfigure}[b]{0.5\linewidth}
        \flushright
     \includegraphics[width=1.1\linewidth]{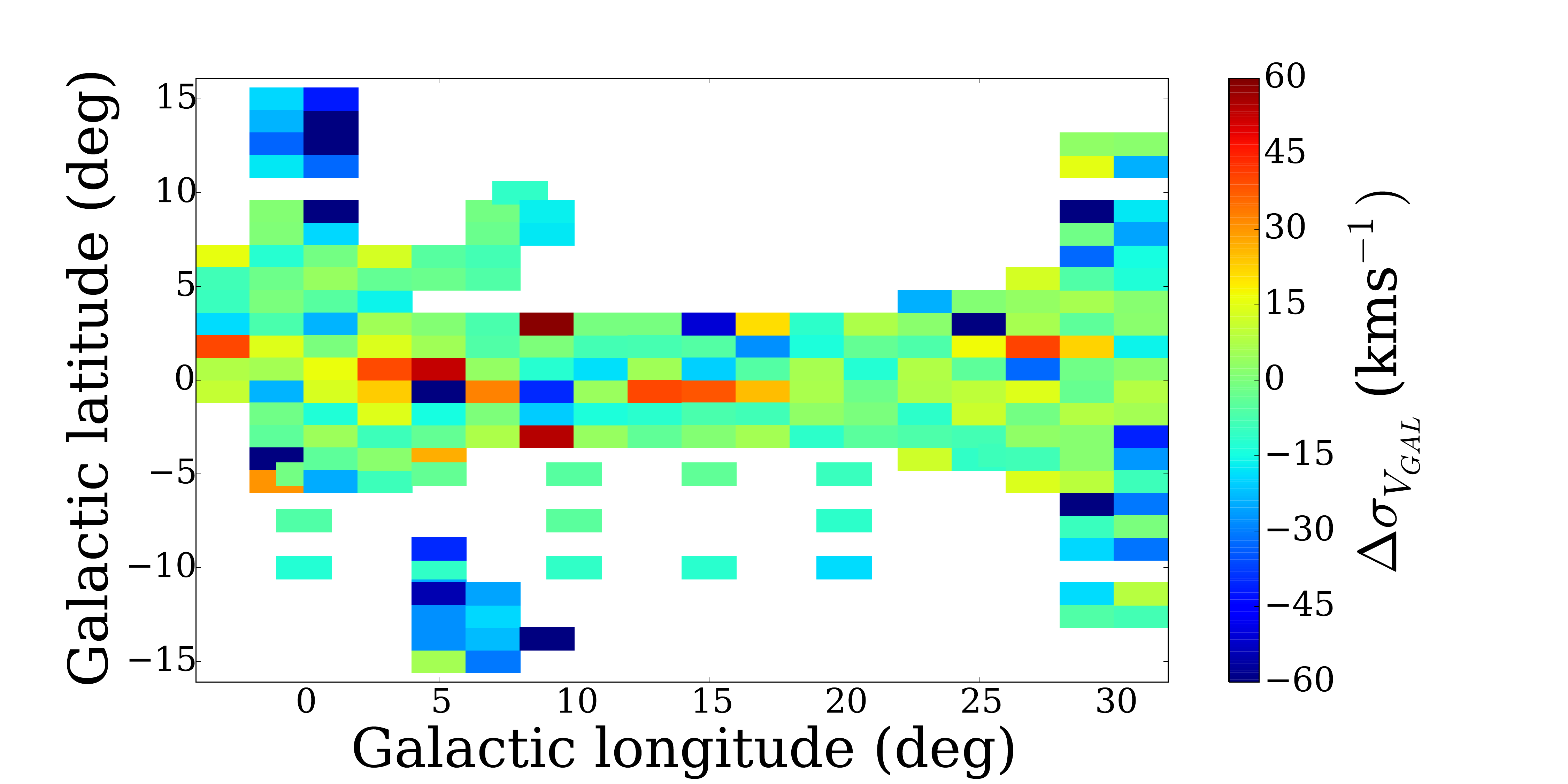}
    \caption{}
  \end{subfigure}%
  \begin{subfigure}[b]{0.5\linewidth}
    \flushright
    \includegraphics[width=1.1\linewidth]{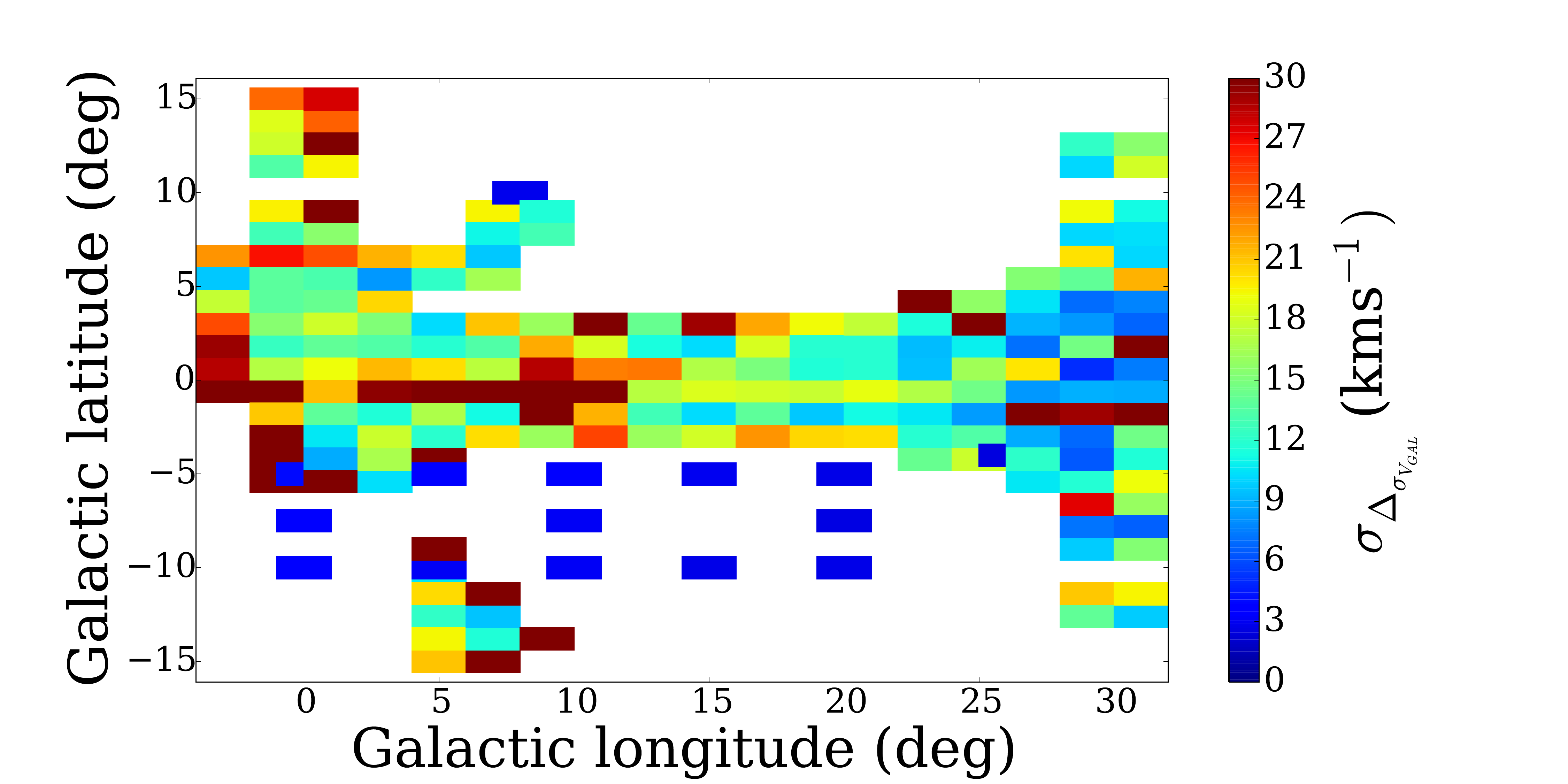}
    \caption{}
     \end{subfigure}
  \caption{Same as Figure~\ref{fig:model1}, but for velocity dispersion. The model is kinematically hotter than the data in the very centre of the bulge as also seen in Figure \ref{fig:ap_ar}. {Overall the model provides a good match to the data across all $(l,b).$}}
  \label{fig:model2}
\end{figure*}

Mean velocity and velocity dispersion maps, comparing data and the model, for \apogee, BRAVA, and ARGOS, across the bulge and moving into the disk, are shown in Figures~\ref{fig:model1} and \ref{fig:model2}, respectively. These Figures include maps of the residuals between data and model and the errors on these residuals, for the ARGOS and \apogee\ fields.  Panels (a) -- (d) include stars only within distances 4 -- 12 kpc from the Sun (except for the BRAVA data where no stellar parameters or subsequently distances were determined for the stars). Panels (a) and (b) show the rotation of the data and model and the dispersion of the data and model, in Figures \ref{fig:model1} and \ref{fig:model2} respectively. Panels (c) and (d) then show the difference between the model and data (model -- data) and the error on that value, respectively.

From Figure \ref{fig:model1}, panels (a) and (b) it is apparent that the overall mean velocity of the {model is a good match to the data}. The residuals in panel (c) do not show coherent angular structure. There are a few mismatching bins between data and model but with correspondingly large errors due to the small numbers of stars along these lines of sight (see panel (d)). Note in Figure \ref{fig:model1}d, the errors are relatively low for the ARGOS fields as there are about 600 stars that comprise these fields, in comparison to $>$ 30 on average (but in some cases lower) in the binned \apogee\ fields comprising the rest of the map.

In the dispersion profiles shown in Figure \ref{fig:model2}, {and as also seen in the bottom panel of Figure \ref{fig:ap_ar}, this particular model is slightly kinematically hotter than these data}, within $|l|< 5^\circ$. As noted by \citet{Ness2013b}, the model is more centrally concentrated than the data. With the exception of the few bins in the mid-plane distributed across a narrow range in latitude, there is no overall structure in the difference in dispersion between model and data, as seen for the rotation. Some {bins in the mid-plane are}, however, significantly kinematically cooler than the model. This is seen just below the plane, for a narrow latitude range ($b$ $\sim$ 1$^\circ$) and this extends from $l$ = 12$^\circ$ out into the disk. {There are very few stars along the line of sight that are located within this apparently kinematically cool structure in the disk, although the sampling errors shown in Figure \ref{fig:model2}d are small and indicate these differences are statistically significant}. 

\begin{figure}[h!]
%\centering
\flushright
    \includegraphics[scale=0.2]{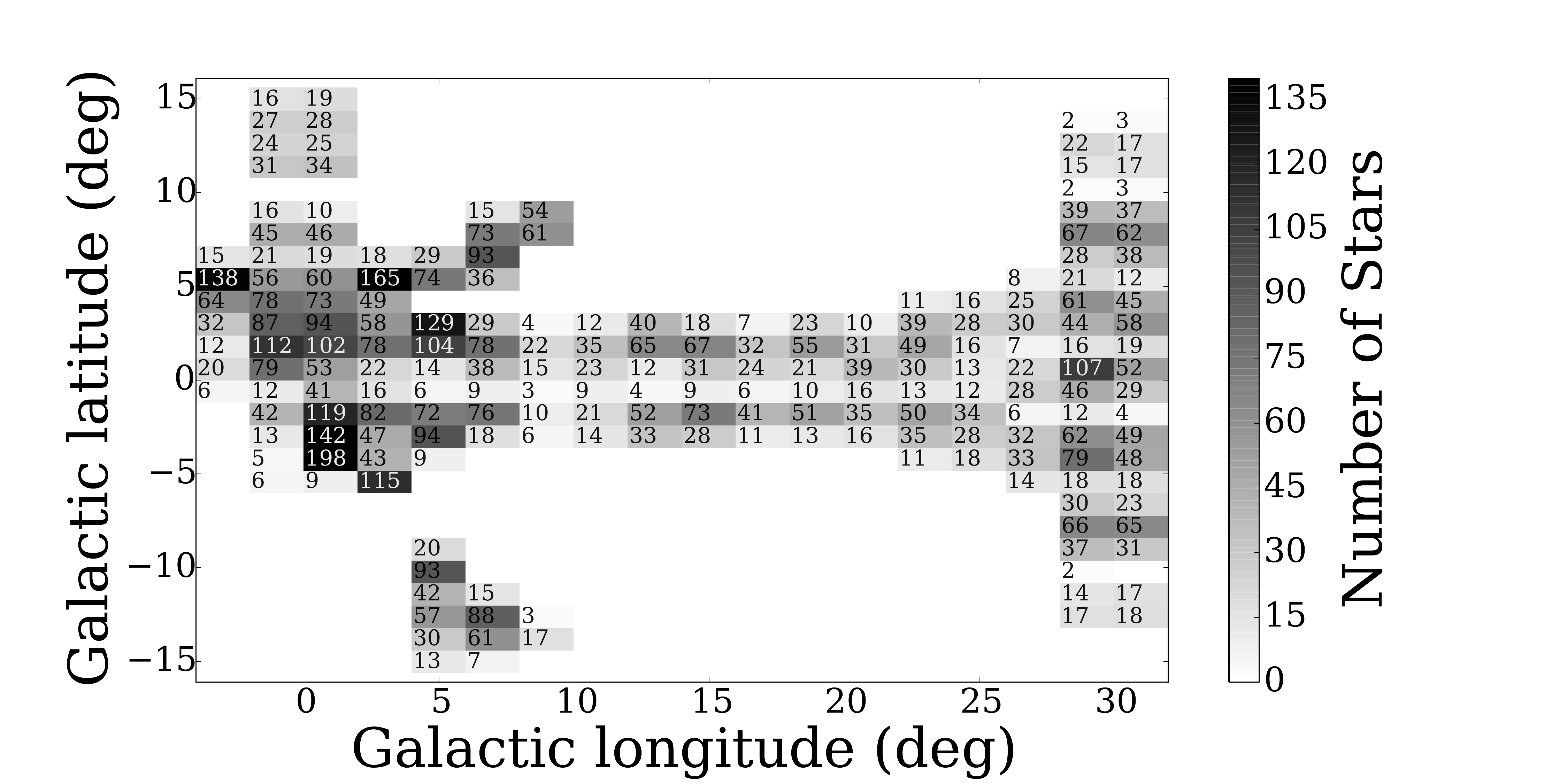}
\caption{The numbers of stars used in Figures \ref{fig:model1} and \ref{fig:model2}. There are a total of $\approx$ 7500 \apogee\ stars in this $(l,b)$ range. There are fewest stars in the plane.}
\label{fig:numsdist}
\end{figure}

{Figure \ref{fig:numsdist} shows the numbers of stars in the fields in Figures \ref{fig:model1}a and \ref{fig:model2}a. {Figure 2} shows that the stars within $|b|$ $<$ 2$^\circ$ are relatively nearer to the Sun compared to stars farther from the plane. Although Figure \ref{fig:model2}d indicates that the low dispersion in some bins near the plane in the data compared to the model is statistically significant, this lower velocity dispersion that is measured may simply reflect the limited extent in distance of these observed stars compared to the surrounding stars. }
The narrow distance dispersion of the stars may also be a consequence of the stellar density, which is expected to peak at the bar, (which is located at about 4.7 kpc away at 30$^\circ$). It is possible that this kinematically cool component is associated with the ``thin" bar, or ``super-thin'' bar structure reported by \citet{Wegg2015} (see Section 5.1). These data may alternatively, reflect the presence of a very cool part of the disk that extends from the outer region into the bulge (rather than being a ``thin" bar). 

Overall, the kinematic behavior of the APOGEE bulge stars is in good agreement with the behavior in the previous fields mapped by ARGOS and BRAVA, and with N-body models of boxy/peanut bulges formed from bars via instabilities of the Galactic disk \citep[see also][]{Shen2010, Portail2015a}.

\section{Chemodynamical signatures of the bulge components}

The kinematics of stars in the bulge is a function of the \feh, and the rotation and dispersion profiles of the stars change with metallicity \citep{Babusiaux2010, Ness2013b}. The \apogee\ data allow examination of the kinematics not only as a function of \feh, for the previously unexamined mid-plane and low-latitude regions, but also \alphafe\ and [X/Fe]. In this first chemodynamical analysis, we report the global trends with \feh.

\subsection{The long bar}

The boxy/peanut extent of the bulge, detected in the COBE/DIRBE image \citep{Dwek1995}, extends to a latitude above the plane of around $b$ $\approx$ 12$^\circ$ and a longitude (on the near side) of about $l$ $\sim$ 15$^\circ$. The boxy/peanut region of the bulge grows out of a thinner bar --- in the innermost boxy region, the stars that were originally in this bar are on orbits now extended in the vertical direction \citep[e.g.,][]{Debattista2005, Inma2006, Athanassoula2005}. 

As expected \citep{Athanassoula2005}, this thinner part of the bar reaches further in the plane of the Galaxy than the $l$ = 15$^\circ$ boxy extent. \citet{Wegg2015} examined the red clump density distribution of stars toward the bulge using VVV photometry and report a $|z| \lesssim 180$~pc thin bar and $|z| \lesssim 45$~pc super-thin bar. The latter of these extends from the centre of the bulge to longitudes $l$ $\approx$ 25-30$^\circ$. The thinnest component has the highest density at the end of the bar, $l \sim$ 30$^\circ$. The transition between the super-thin bar and thin bar structures occurs at a latitude of $|b| \sim 2^\circ$. Figure \ref{fig:model2} demonstrated the presence of a kinematic structure not seen in the N-body model, which, given its limited extent in $b$ is more likely associated with the super-thin bar than the thin bar, but may also be simply part of the disk that is being observed across a narrow range in distance. We now look for signatures of this thin bar in the \apogee\ data that might correlate with the morphology, \feh, and \alphafe.

{The mean \feh\ and \alphafe\ of stars in the bulge and inner disk are shown in Figures \ref{fig:metals1} and \ref{fig:metals2}. 
The trends revealed in these maps are discussed in detail in A.~E.~Garc\'ia-P\'erez et al., (2016, in prep). These trends are sensitive to the derived distances and subsequent distance cuts and selection of stars included in the analysis. Care must be taken comparing the mean metallicity across longitude, as the \apogee\ target selection is different for bulge (357$^\circ$ $\le$ $l$ $\le$ 22$^\circ$) and disk \citep[see][]{Zasowski2013}. Brighter stars are targeted in the bulge and the bulge metallicity distribution contains more cooler, lower \logg\ stars, compared to the disk sample. Figures \ref{fig:metals1}a and \ref{fig:metals2}a show the \feh\ and \alphafe\ maps made for all 10,000 stars at distances 4 - 12 kpc. The majority of foreground stars, which are shown in Figure \ref{fig:metalsnear} for a distance selection of $<$ 3 kpc, are metal-rich stars; these metal-rich foreground stars are eliminated from the bulge maps by performing these distance cuts.  }

{For comparison, Figures \ref{fig:metals1}b and \ref{fig:metals2}b show the sample of these stars with the same distance cuts, but with a lower \logg\ limit imposed of $\logg$ $>$ 0.5 dex. This removes a large number of the lowest \logg\ stars which are targeted in larger fraction in the bulge compared to the disk, due to the brighter magnitude limit of the bulge selection strategy.  This effectively eliminates many of the more distant, low metallicity stars that have higher relative fractions within $l$ $<$ 22$^\circ$. This demonstrates how the metallicity map across longitude is sensitive to distance biases. Consistent qualitative trends however, illustrated in these Figures \ref{fig:metals1}a and and  \ref{fig:metals1}b and \ref{fig:metals2}a and  \ref{fig:metals2}b are, that the highest metallicity stars, with corresponding lowest \alphafe\ enhancements, are concentrated to within $|b|$ $\lesssim$ 2$^\circ$. The apparent metal-rich structure in the mid-plane ($b < 2^\circ$), with a corresponding low $\alpha$-enhancement (as seen in Figure \ref{fig:metals2}), may be associated with the thin bar of \citet{Wegg2015}. }

Figures \ref{fig:metals1} and \ref{fig:metals2} show that there is a homogeneity in the mean \feh\ and \alphafe\ within $(l,b)$ $<$ (5$^\circ$,5$^\circ$), the inner-most region of the boxy part of the bar. There is a known metallicity gradient with latitude in the bulge, for latitudes $|b| > 5^\circ$, {of about -0.45 dex/kpc \citep[e.g.][]{zoccali2008, Ness2013a, RJ2014} and a smaller gradient in longitude \citep[e.g.][]{Gonzalez2013}}. The flattening of the metallicity gradient with latitude near the plane has been detected \citep{ramirez2000, rich2007, rich2012} and the \apogee\ survey now demonstrates that flattening is present not only at the minor axis but the entire extent of the boxy bulge ($l$ $\lesssim$ 15$^\circ$) within $|b|$ $<$ 2$^\circ$.

There is a relatively sudden decrease in \feh\ at $|b|$ $>$ 5$^\circ$ above the plane, seen in Figures \ref{fig:metals1} and \ref{fig:metals2} across $l$.  There is also a sudden increase in \alphafe\ for all latitudes above $b$ $>$ 2$^\circ$ as seen in Figure \ref{fig:metals2}. The metal-rich stars concentrated in the plane bar may reflect that there is ongoing star formation and even a potentially younger population in the plane extending out into the disk \citep{Ness2014}.

\subsection{The long bar and boxy bulge}

\begin{figure*}
    \begin{subfigure}[b]{0.5\textwidth}
    \centering
        \includegraphics[width=1.1\linewidth]{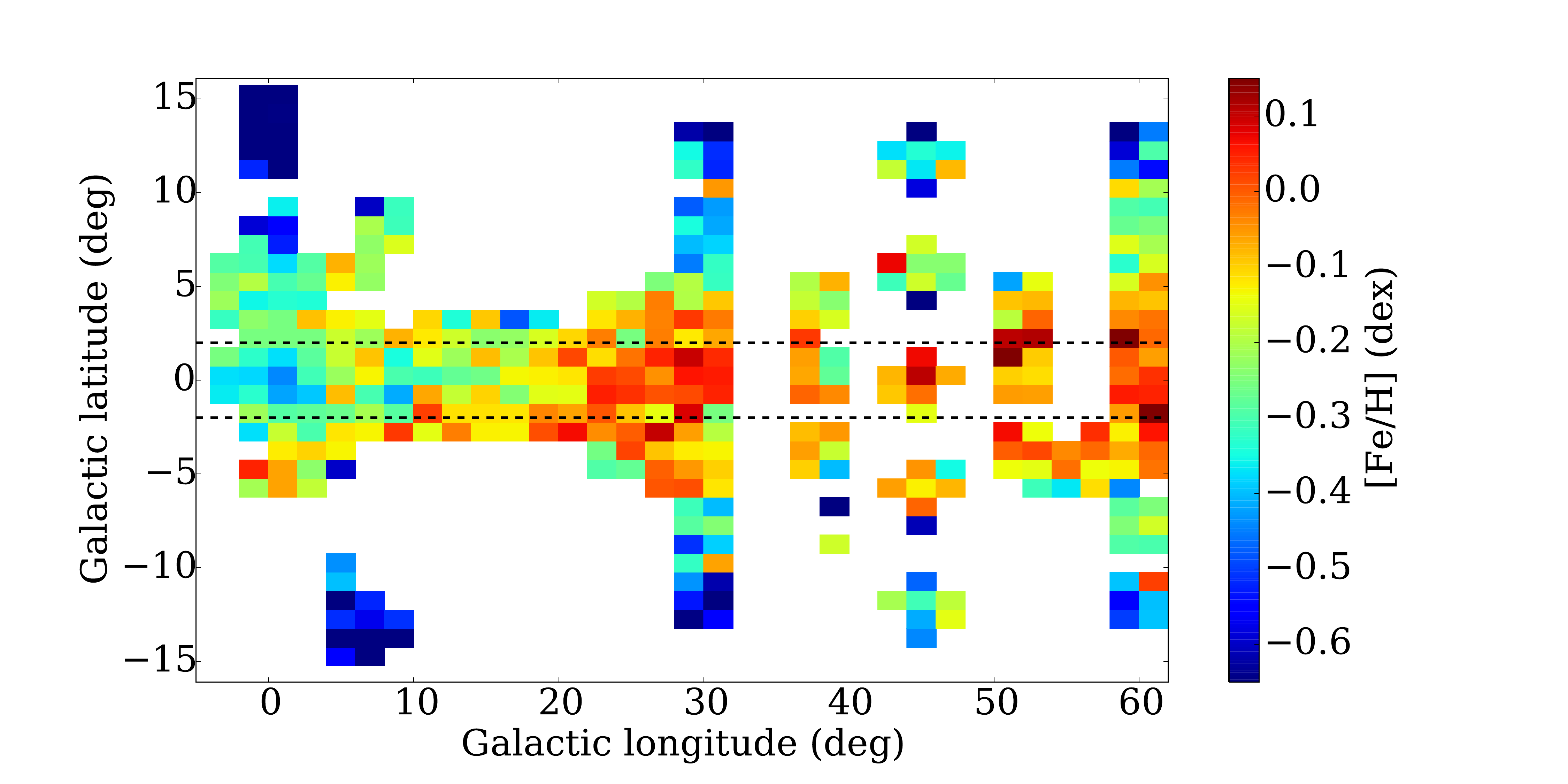}
\caption{}
  \end{subfigure}
    \begin{subfigure}[b]{0.5\textwidth}
    \centering
   \includegraphics[width=1.1\linewidth]{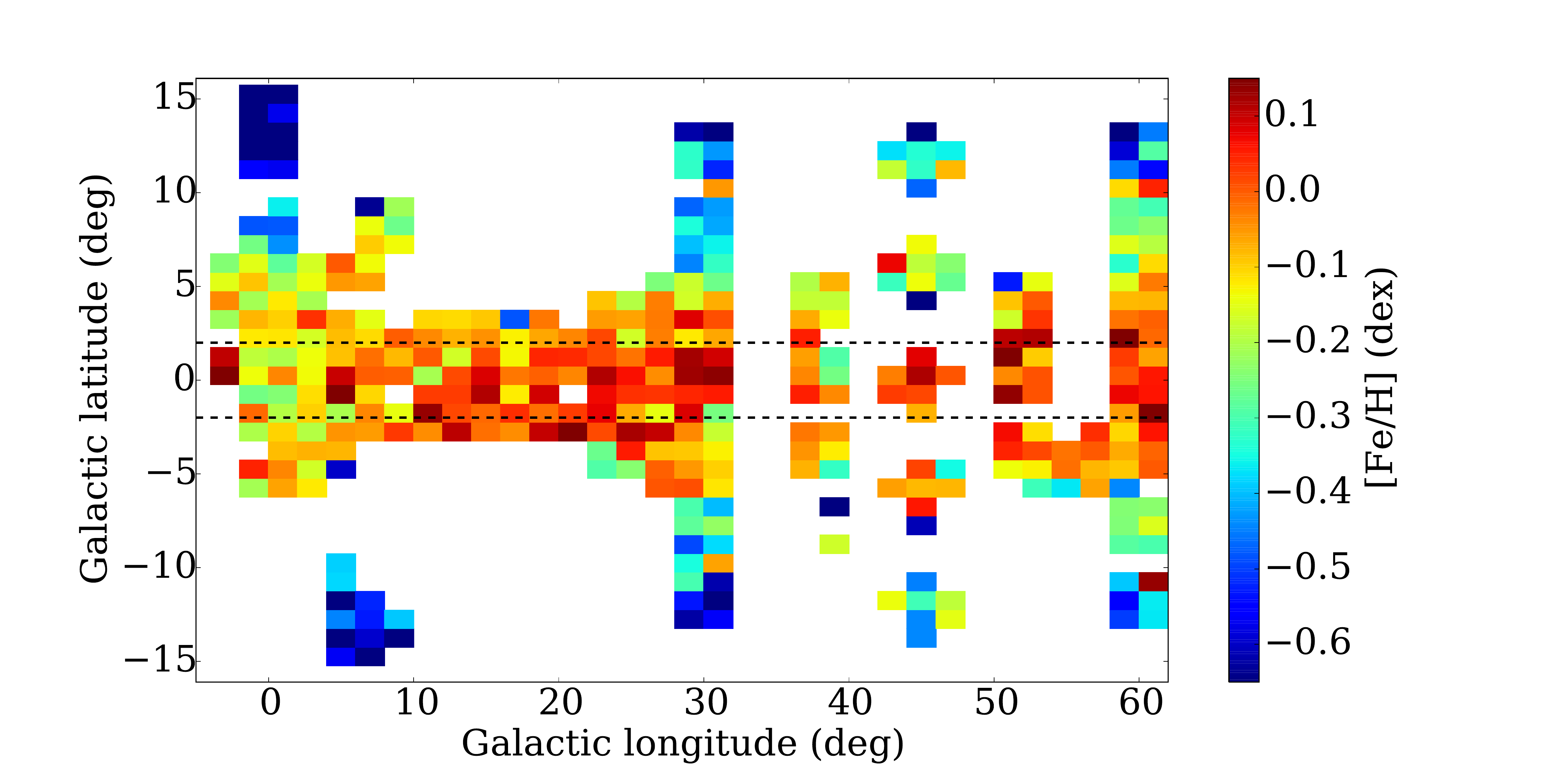}
\caption{}
  \end{subfigure}%
  \caption{ \feh\ maps for the (a) 10,000 bulge and disk stars and (b) 8,000 bulge with \logg\ cuts \logg\ $>$ 0.5 dex eliminating the most distant metal-poor stars from the bulge sample, where they are represented in higher number than the disk; both plots span heliocentric distances of 4--12  kpc, showing the mean \feh\ values in each bin. The dashed line indicates the scale height of the 180~pc thin bar identified by \citet{Wegg2015}. The highest-metallicity stars are found within $b$ $<$ 2$^\circ$.}
   \label{fig:metals1}
  \end{figure*}

\begin{figure*}
    \begin{subfigure}[b]{0.5\textwidth}
    \centering
        \includegraphics[width=1.1\linewidth]{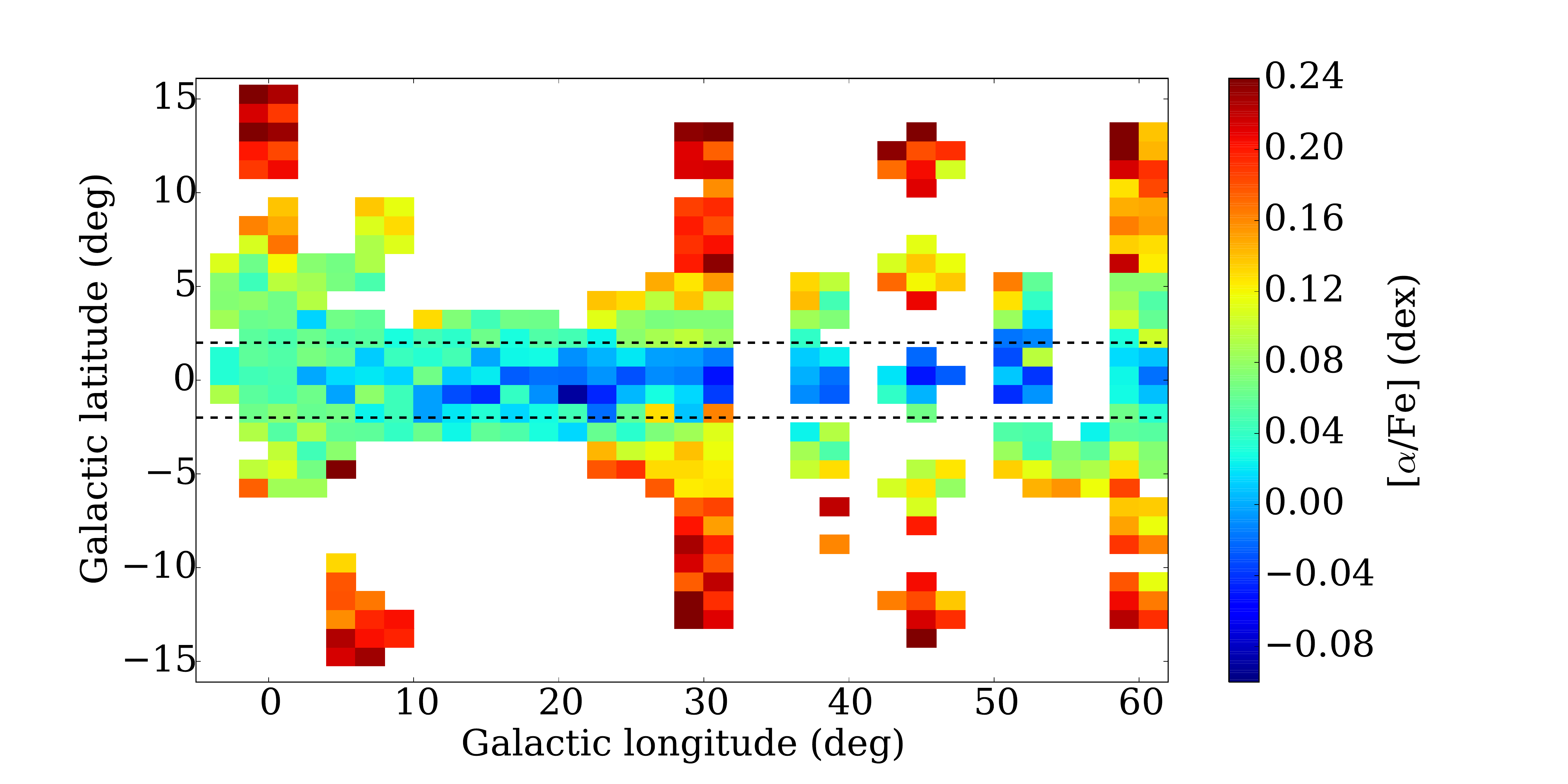}
\caption{}
  \end{subfigure}
    \begin{subfigure}[b]{0.5\textwidth}
    \centering
   \includegraphics[width=1.1\linewidth]{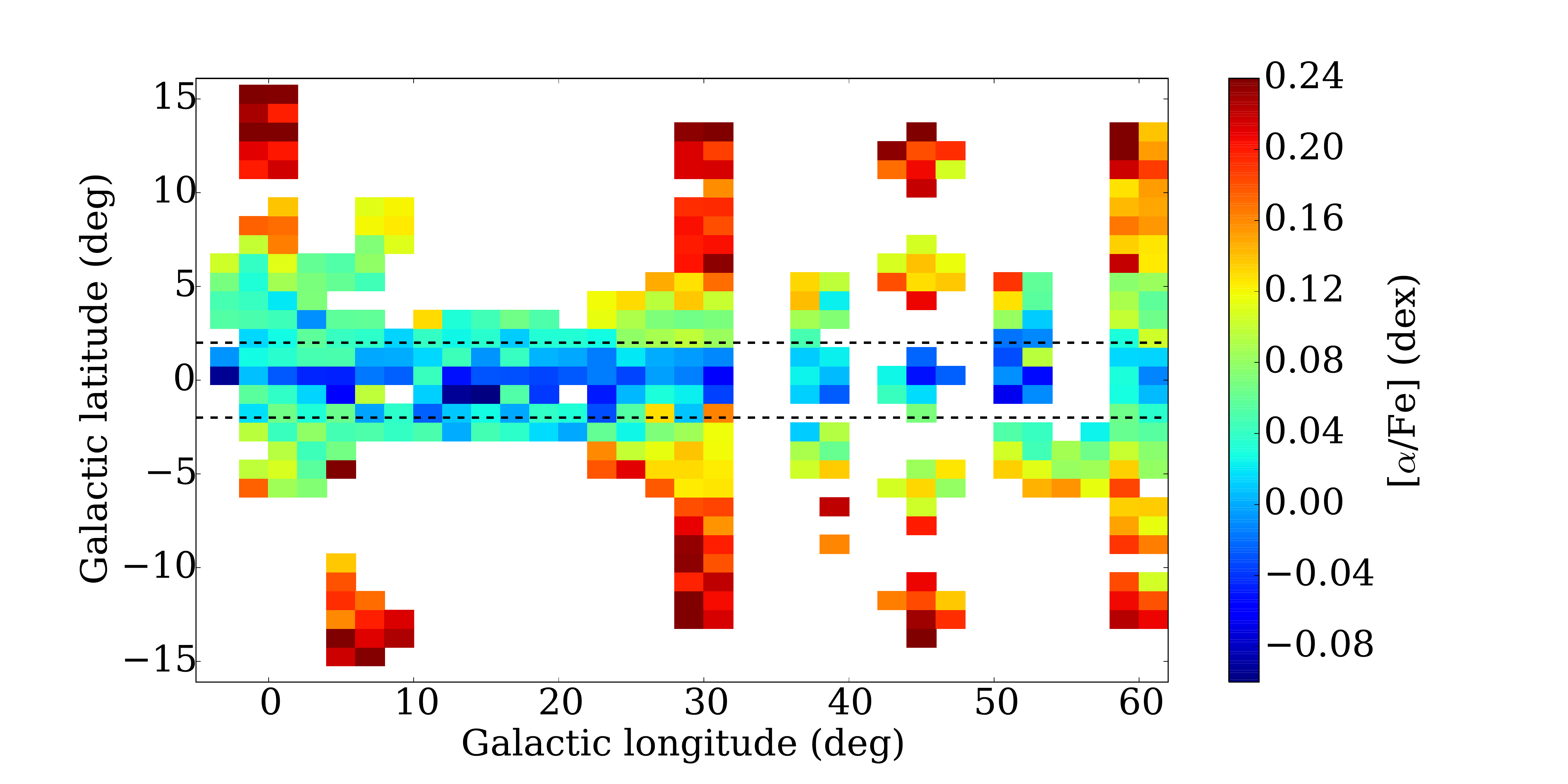}
\caption{}
  \end{subfigure}%
  \caption{ Similar to Figure \ref{fig:metals1} but for \alphafe. The lowest \alphafe\ stars are found within $b$ $<$ 2$^\circ$.}
   \label{fig:metals2}
  \end{figure*}

\begin{figure*}
    \begin{subfigure}[b]{0.5\textwidth}
    \centering
        \includegraphics[width=1.1\linewidth]{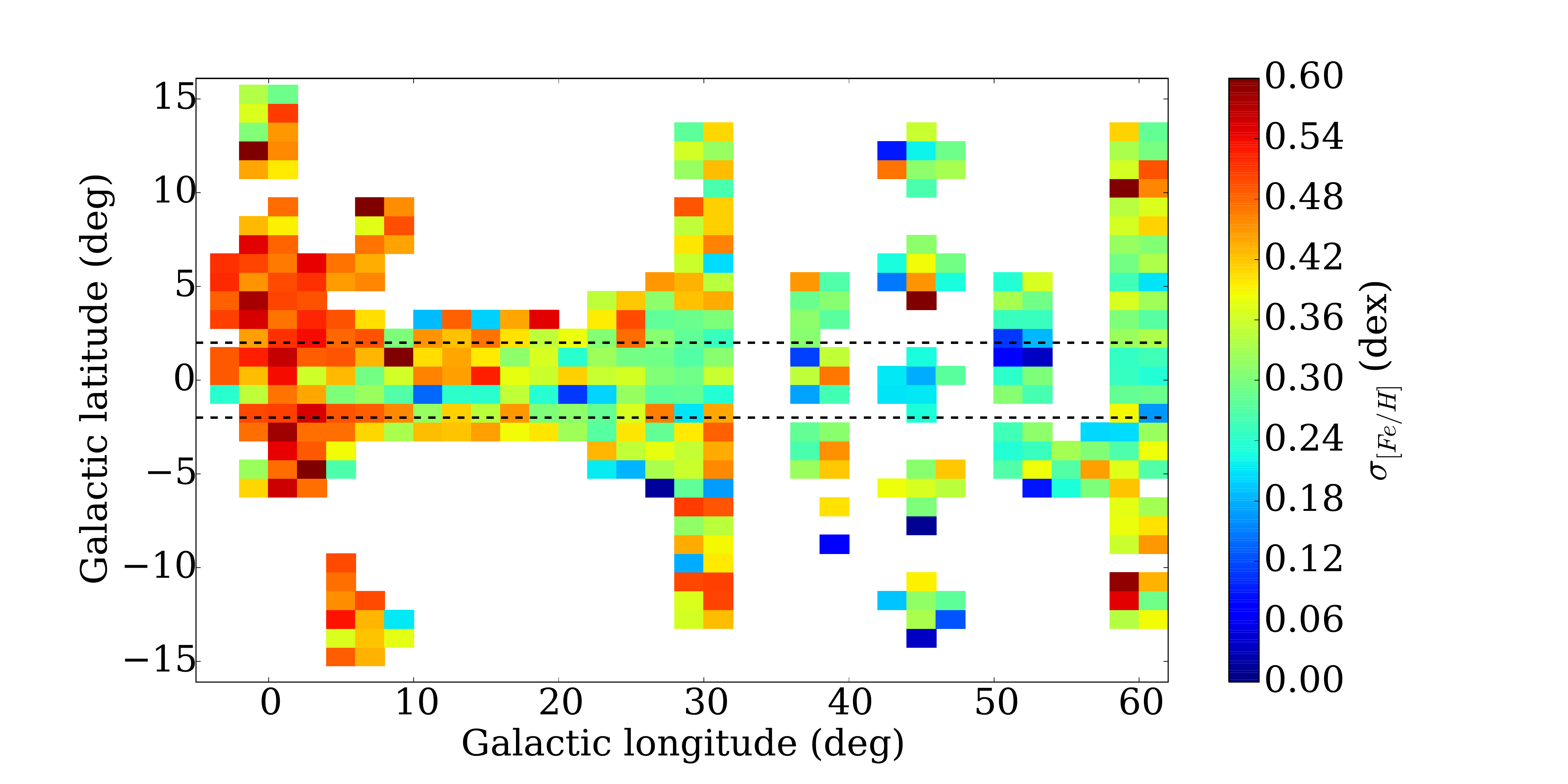}
\caption{}
  \end{subfigure}
    \begin{subfigure}[b]{0.5\textwidth}
    \centering
   \includegraphics[width=1.1\linewidth]{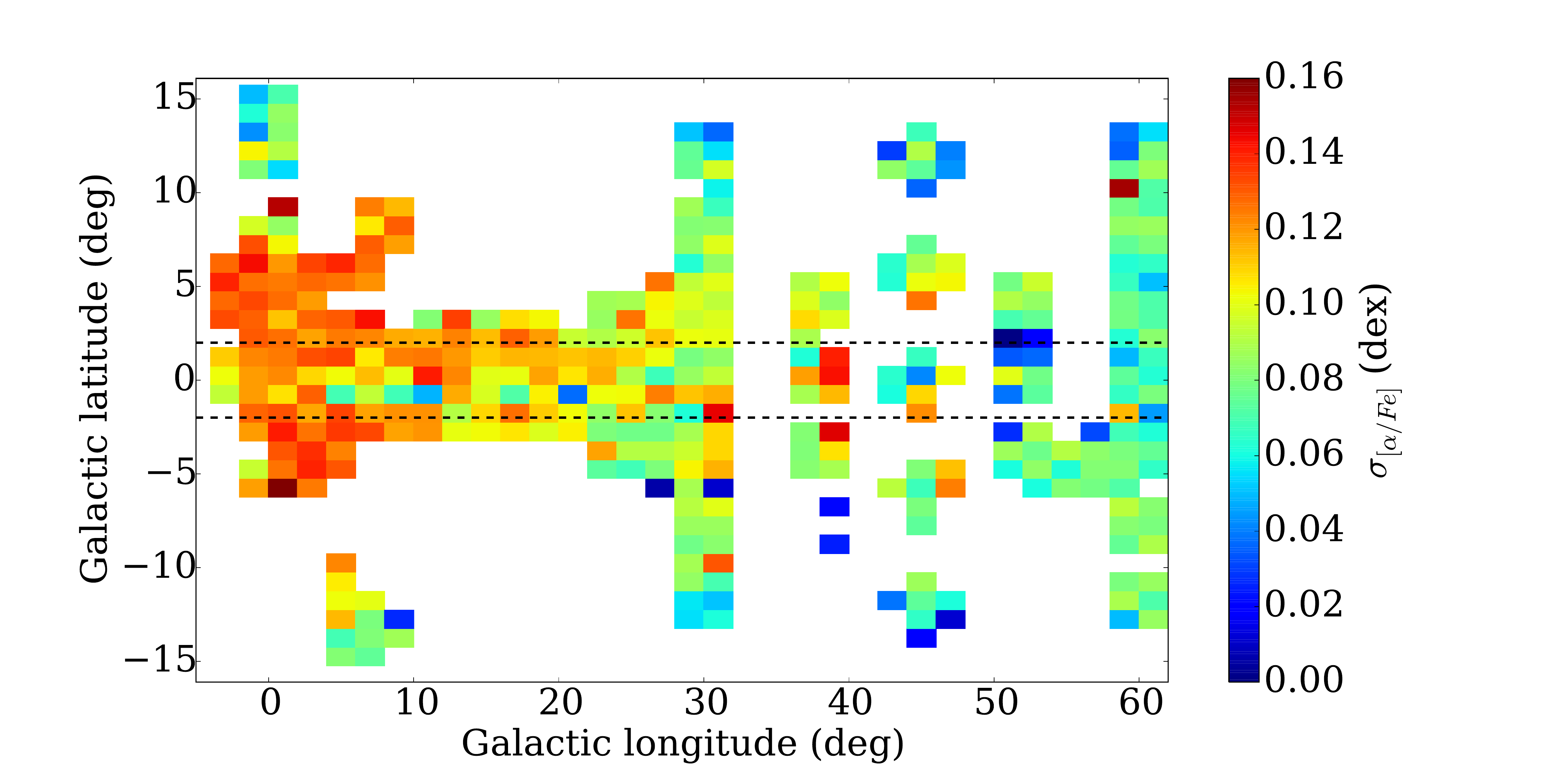}
\caption{}
  \end{subfigure}%
    \caption{ Dispersion maps of (a) \feh\ and (b) \alphafe,  for the 10,000 bulge and disk stars spanning heliocentric distances of 4--12  kpc. }
\label{fig:metals3}
  \end{figure*}

{Figure \ref{fig:metals3} shows the dispersion in \feh\ and \alphafe\ of the bulge and disk. These trends in \feh\ and \alphafe\ dispersion are very similar for both selections compared in Figures \ref{fig:metals1}a and  \ref{fig:metals1}b and \ref{fig:metals2}a and  \ref{fig:metals2}b, so only the full selection in (a) of these Figures is {shown}. Figure \ref{fig:metals3} illustrates that the MDF is broadest at the minor axis.  As seen in Figure \ref{fig:metals3}a, the \feh\ dispersion is largest on the minor axis, at all $b$ and as seen in Figure \ref{fig:metals3}b, the \alphafe\ dispersion is highest within $(l,b)$ $\lesssim$ (5$^\circ$, 5$^\circ$).  The broadest MDF in the inner most region may reflect the mixture of populations in the centre of the Milky Way \citep[e.g.][]{Ness2013a, Gonzalez2015}. {The population at high latitudes is metal-poor and alpha-rich (see Figure \ref{fig:metals1} and \ref{fig:metals2} and, at high latitudes, both disk and bulge have very similar mean metallicity, (\feh, \alphafe) and metallicity dispersion, $\sigma$(\feh, \alphafe)}. This suggests a homogeneous, likely  thick disk population extending to the inner most region of the Milky Way, similar to the thicker part of the disk seen near the Sun \citep{Ness2013a}. Correspondingly, the kinematic behavior is smooth moving from inner to outer region, as seen from Figure \ref{fig:rotstd}.}

The metallicity dispersion {in the bulge} shown in Figure \ref{fig:metals3} is $>$ 50\% lower outside of $l$ $\gtrsim$ 10$^\circ$ compared to inside of $l$ $\lesssim$ 10$^\circ$. The lowest dispersion in \feh\ is in the plane in the disk at $l$ $>$ 30$^\circ$, within $|b|$ $<$ 2$^\circ$ (Figure \ref{fig:metals3}a). The dispersion in \alphafe\ decreases by about 50\% outside of (l,b) $<$ (10$^\circ$,5$^\circ$) (Figure \ref{fig:metals3}b) and also decreases above $b$ $>$ 10$^\circ$ at the minor axis.

Figure \ref{fig:rotstd}a shows that the overall trends in the velocity are smooth, with no obvious line-of-sight velocity signatures associated with the changing gradients in metallicity across $(l,b)$ seen in Figure \ref{fig:metals1}.  In Figure \ref{fig:rotstd}b, we see that the bulge is kinematically hottest in the inner-most region $(l,b)$ $<$ (5$^\circ$,5$^\circ$), and the dispersion drops off rapidly in $b$ outside of $b$ $>$ 5$^\circ$; this pattern corresponds to where the mean metallicity drops in latitude, for longitudes within $l$ $<$5$^\circ$. Otherwise, at lower latitudes, the metallicity within $l$ $<$ 5$^\circ$ is fairly constant. Outside of longitudes $l$ $>$ 10$^\circ$, the dispersion, shown in Figure \ref{fig:rotstd}b, decreases gradually into the disk and becomes nearly constant with latitude as longitude increases. 

\begin{figure*}
    \begin{subfigure}[b]{0.5\textwidth}
    \centering
        \includegraphics[width=1.1\linewidth]{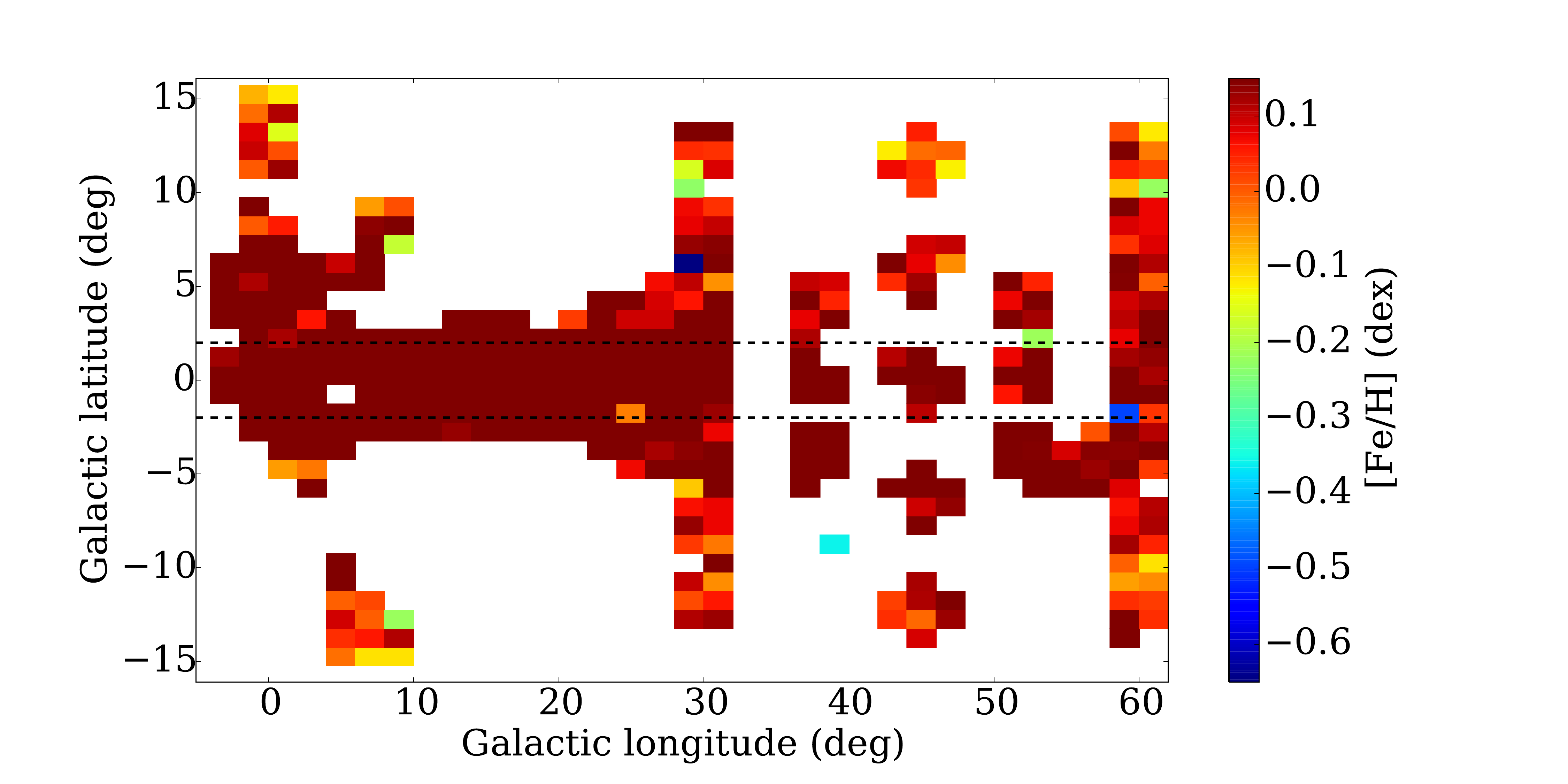}
\caption{}
  \end{subfigure}
    \begin{subfigure}[b]{0.5\textwidth}
    \centering
   \includegraphics[width=1.1\linewidth]{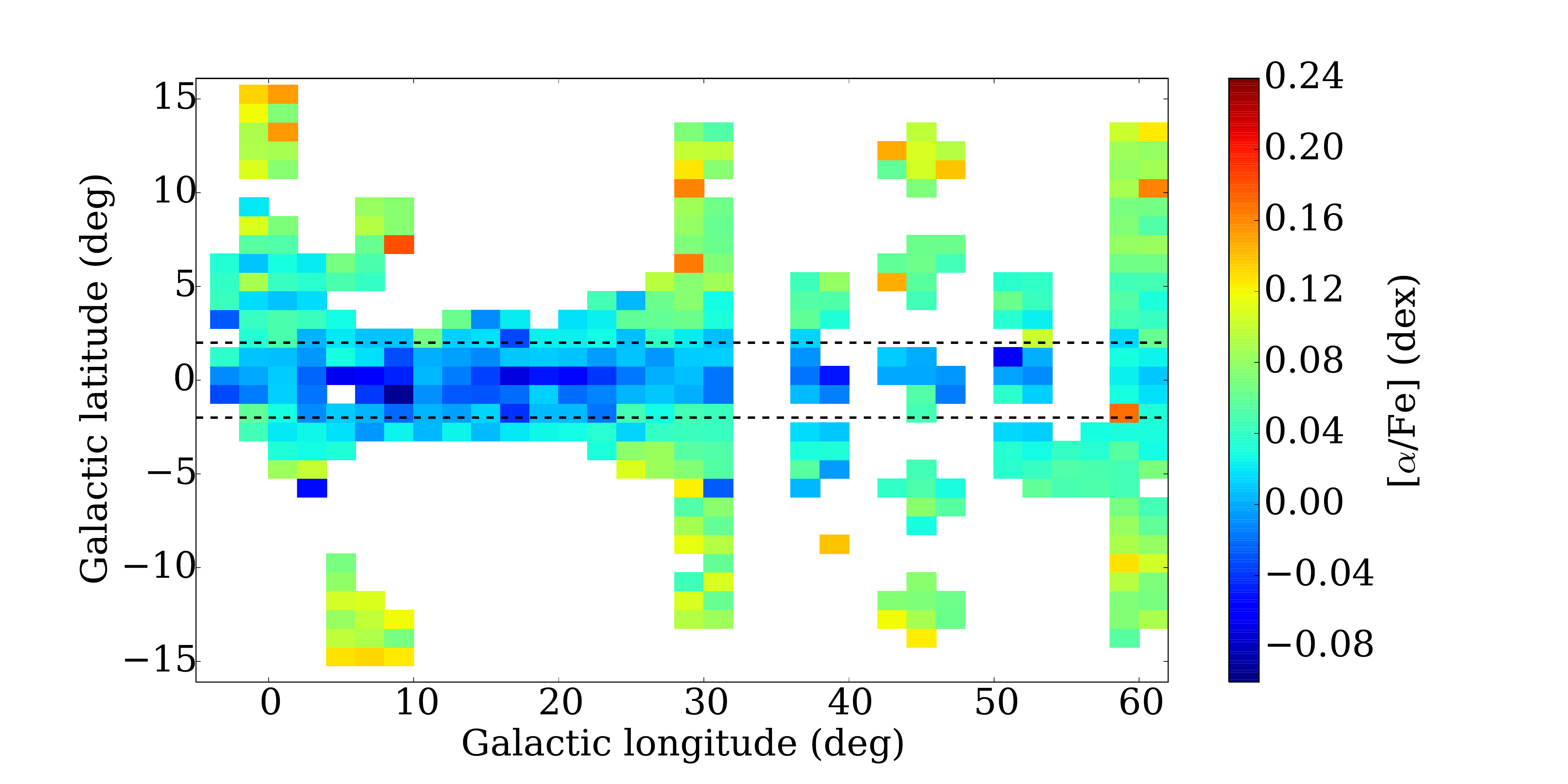}
\caption{}
  \end{subfigure}%
 \caption{ Similar to Figure~\ref{fig:metals1}, but for 6800 \apogee\ foreground disk stars eliminated from the bulge sample, with heliocentric distances of $<$3~kpc.  The foreground population is homogeneously metal-rich.}
   \label{fig:metalsnear}
  \end{figure*}

{Figure  \ref{fig:metalsnear} contrasts the \feh\ and \alphafe\ map of stars in the bulge with the foreground disc stars with distances $<$ 3 kpc from the Sun. The foreground disk is homogeneously metal-rich to latitudes $b$ $<$ 10$^\circ$. It is also alpha-poor at  low latitudes, $b$ $<$ 5$^\circ$ and the  $\alpha$-enhancement increases at higher latitudes.}

\subsection{Kinematics as a function of [Fe/H] into the mid-plane} \label{sec:kinematics_feh}

We now examine the \apogee\ bulge stars as a function of \feh, by dividing the stars into the same four metallicity bins used in the ARGOS analysis of the bulge \citep{Ness2013b}: A ($\feh\ > 0$), B ($-0.5 \le \feh\ < 0$), C ($-1.0 \le \feh\ < -0.5$), and D ($\feh\ \le -1.0$).
We choose these relatively coarse bins for this analysis, because whilst the spatial coverage of the fields is extensive in the plane, there are few stars per field. {We refer to these metallicity bins as components or populations but this is not intended to tie them a separate formation origin. Rather, this is intended {as a tool} to understand the changing kinematics of stars as a function of \feh. }

\begin{figure*}
\centering
    \includegraphics[width=0.7\linewidth]{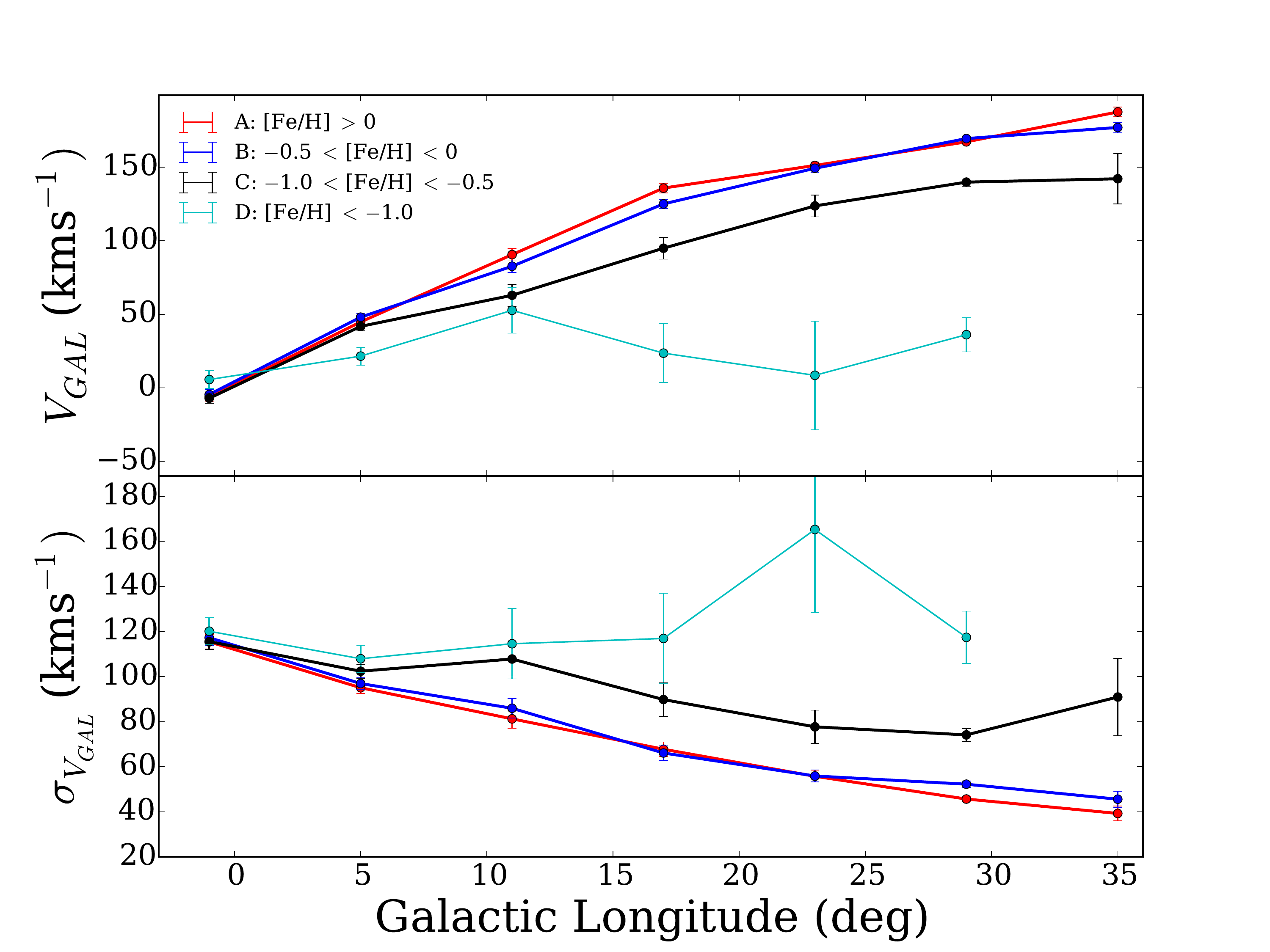}   \\
\label{fig:rotstda}
  \caption{{{Rotation and velocity dispersion profiles for the stars as a function of \feh, combined in latitude and binned in longitude, for metallicity bins A,B,C and D (2700, 2700, 1700 and 450 stars out to $l$ = 35$^\circ$, respectively)}.  {Stars are binned in longitude ranges of $\pm$ 3$^\circ$ around each point shown}. Stars {with} \feh\ $>$ --0.5 have a very similar rotation speed and dispersion profile and stars with  --1.0 $<$ \feh\ $<$ --0.5  have a slower rotation and flatter dispersion across all $l$ than the more metal-rich stars. All stars {with} \feh\ $>$ --1.0 transition smoothly in kinematics from the boxy bulge into the disk. Stars with \feh\ $<$ --1.0 show a changing rotation and dispersion profile outside of the boxy bulge ($l$ $>$ 15$^\circ$). {Inside of the boxy bulge the metal-poor stars, with} \feh\ $<$ --1.0 show the slowest rotation and flattest dispersion.}}
  \label{fig:ABCD3}
\end{figure*}

{Figure  \ref{fig:ABCD3} presents the rotation and dispersion profiles for stars out to $l$ = 35$^\circ$, for all latitudes combined {and across longitude bins of $\pm$ 3$^\circ$}, for each of these four metallicity bins. The rotation is shown in the top panel and the dispersion at bottom. These stars show different kinematics as a function of \feh. Stars in components A and B, with \feh\ $>$ --0.5, show similar rotation and dispersion although the most-metal-rich stars are the kinematically coolest population. Stars in metallicity bin C, with --1.0 $<$ \feh $<$ --0.5,  show a similar, but slower rotation profile to the more metal-rich stars and a hotter and flatter dispersion profile than stars with \feh\ $>$ --0.5 as a function of longitude. All stars with \feh\ $>$ --1.0 (in components A,B,C) show a rotation and dispersion that transitions smoothly from the boxy bulge region inside of $l$ $<$ 15$^\circ$ to the disk.}

 {Stars with \feh\ $>$ --0.5 have a dispersion and rotation profile that is seen in the N-body models of boxy/peanut bulges (shown in Figure \ref{fig:ap_ar}). Stars with \feh\ $<$ --0.5 have a dissimilar dispersion profile than the N-body models. However the rotation and dispersion profile of stars with --1.0 $<$  \feh\ $<$ --0.5 and their smooth transition into the disk suggest these stars belong to the thicker part of the disk in the inner region that is not part of the boxy/peanut morphology \citep[e.g.][]{Ness2013b} and \citet{PdiM2015}.  }
 
{ Stars in metallicity bin D, with \feh\ $<$ --1.0, show the slowest rotation. The rotation and dispersion of these metal-poor stars diverges most significantly from that of the more metal-rich stars outside of the bulge at $l$ $>$ 10$^\circ$. Notably, outside of the boxy bulge region this population of stars {with} \feh\ $<$ --1.0 does not appear to be rotating, whereas inside of $l$ $<$ 15$^\circ$, the stars {with} \feh\ are slowly rotating, on the order of about 50\% of the more metal-rich stars with \feh\ $>$ --0.5. }

{The consistency in the rotation of stars with \feh\ $>$ --1.0 and smooth transition out in longitude suggests these stars have a common origin, {from the disk}, although the more metal-poor stars with \feh\ $<$ --0.5 do not have a dispersion profile that matches N-body models of a boxy bulge formed via instabilities. Stars with \feh\ $<$ -1.0 are likely halo stars in the inner region or perhaps a small population of stars in the bulge that is unique to the bulge (formed via mergers or dissipational collapse). Note that stars with \feh\ $<$ --1.0 represent only 5\% of stars in the bulge, so this corresponds to a very small population.} % of stars with an origin that is not of the disk. }

\begin{figure*}
    \begin{subfigure}[b]{0.5\linewidth}
\flushright
    \includegraphics[width=0.9\linewidth]{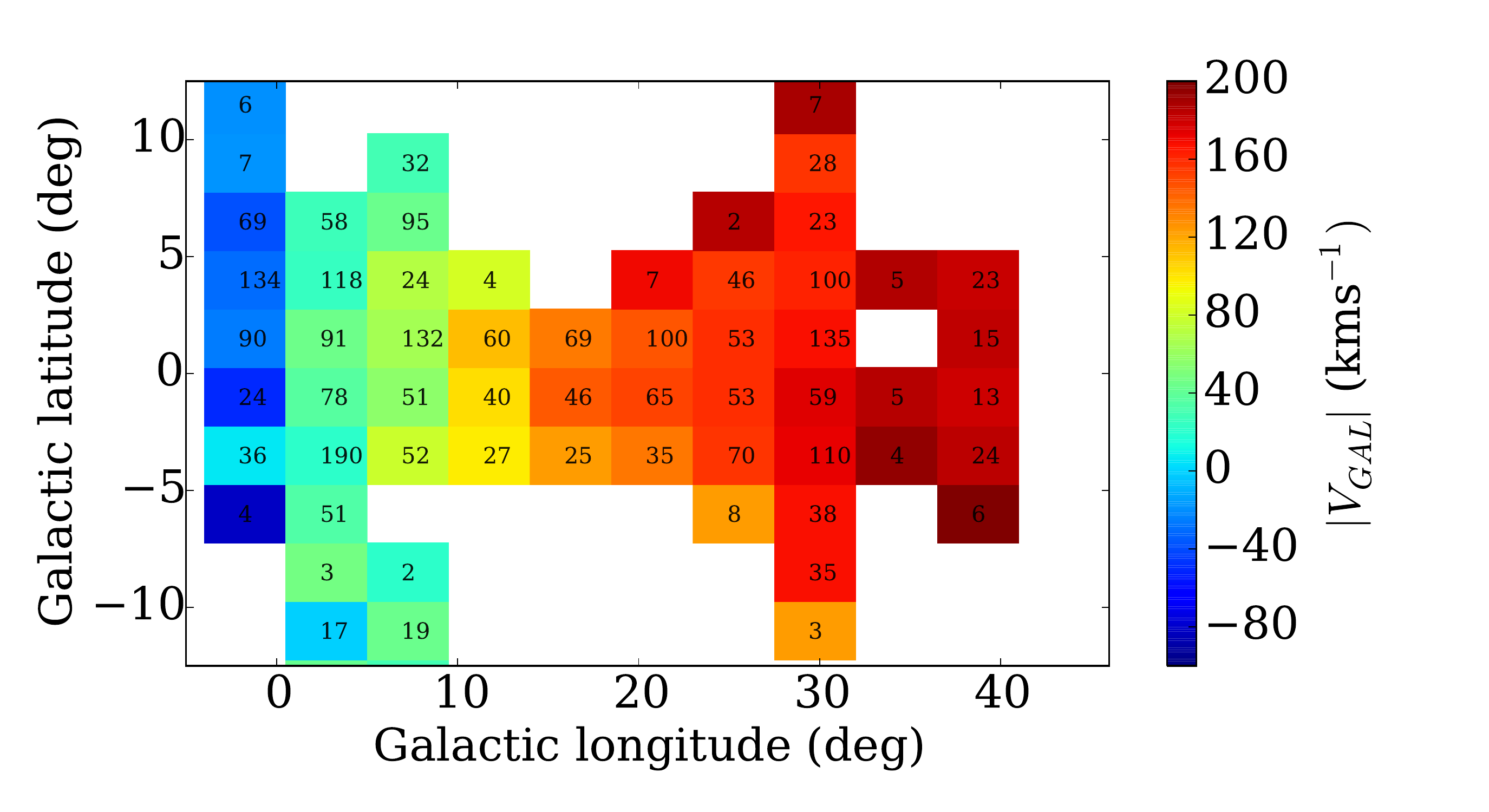}
\caption{A: \feh $>$ 0}
  \end{subfigure}%
  \begin{subfigure}[b]{0.5\linewidth}
\flushright
    \includegraphics[width=0.9\linewidth]{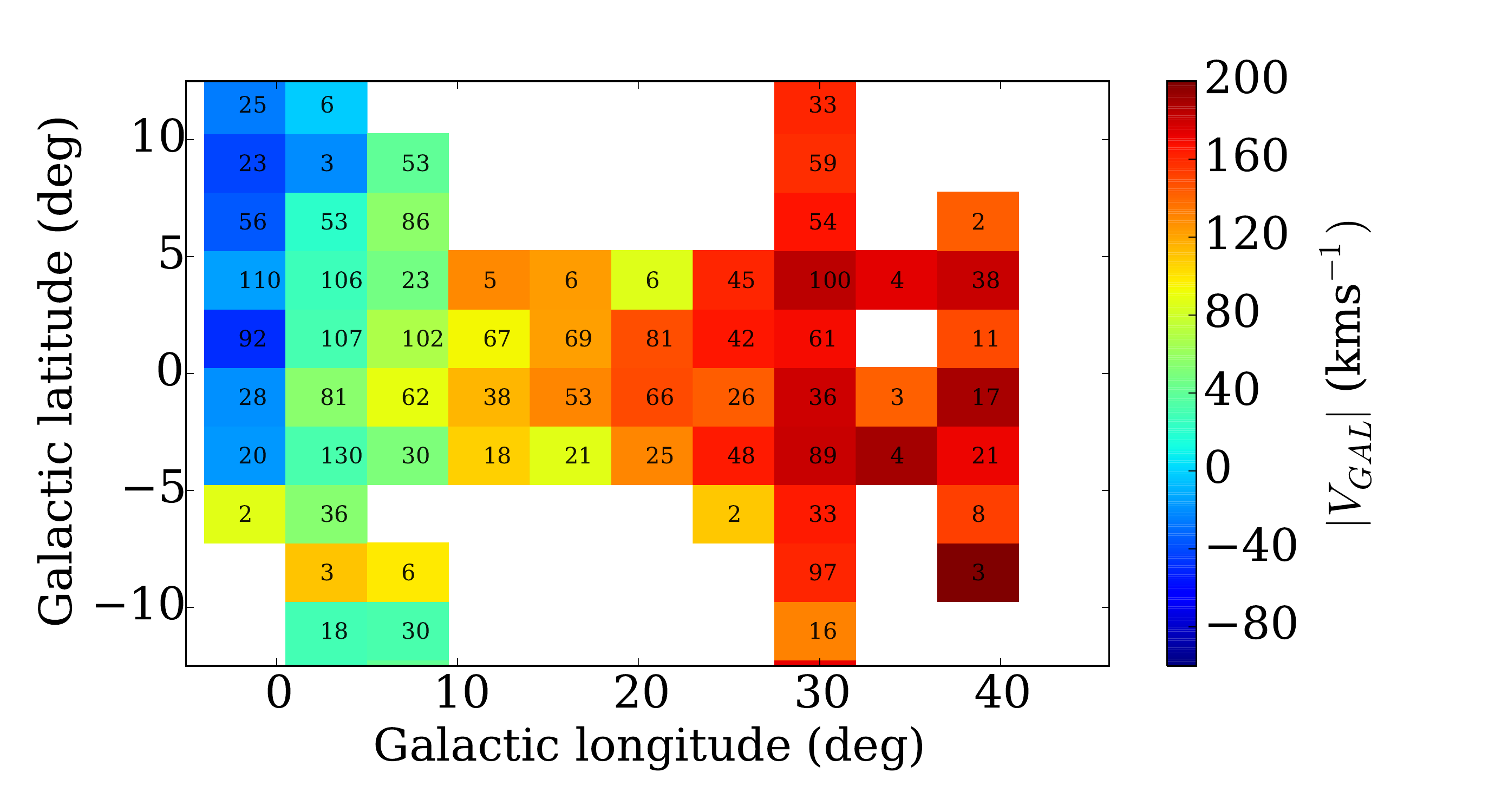} 
\caption{B:   --0.5 $<$ \feh\ $<$ 0 } 
\end{subfigure}
    \begin{subfigure}[b]{0.5\linewidth}
\flushright
    \includegraphics[width=0.9\linewidth]{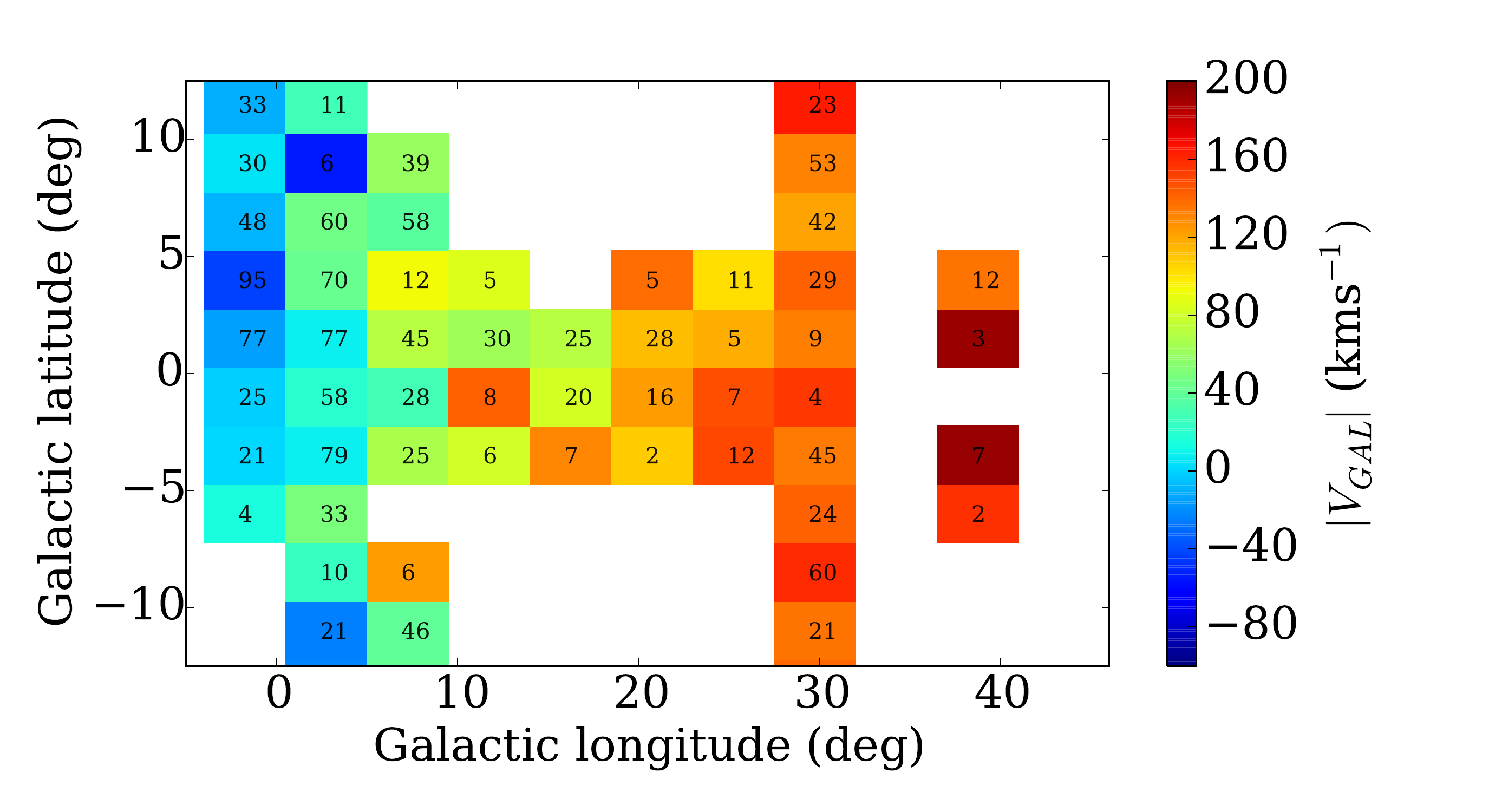}
\caption{C:   --1.0 $<$ \feh\ $<$ --0.5 }
  \end{subfigure}%
  \begin{subfigure}[b]{0.5\linewidth}
\flushright
    \includegraphics[width=0.9\linewidth]{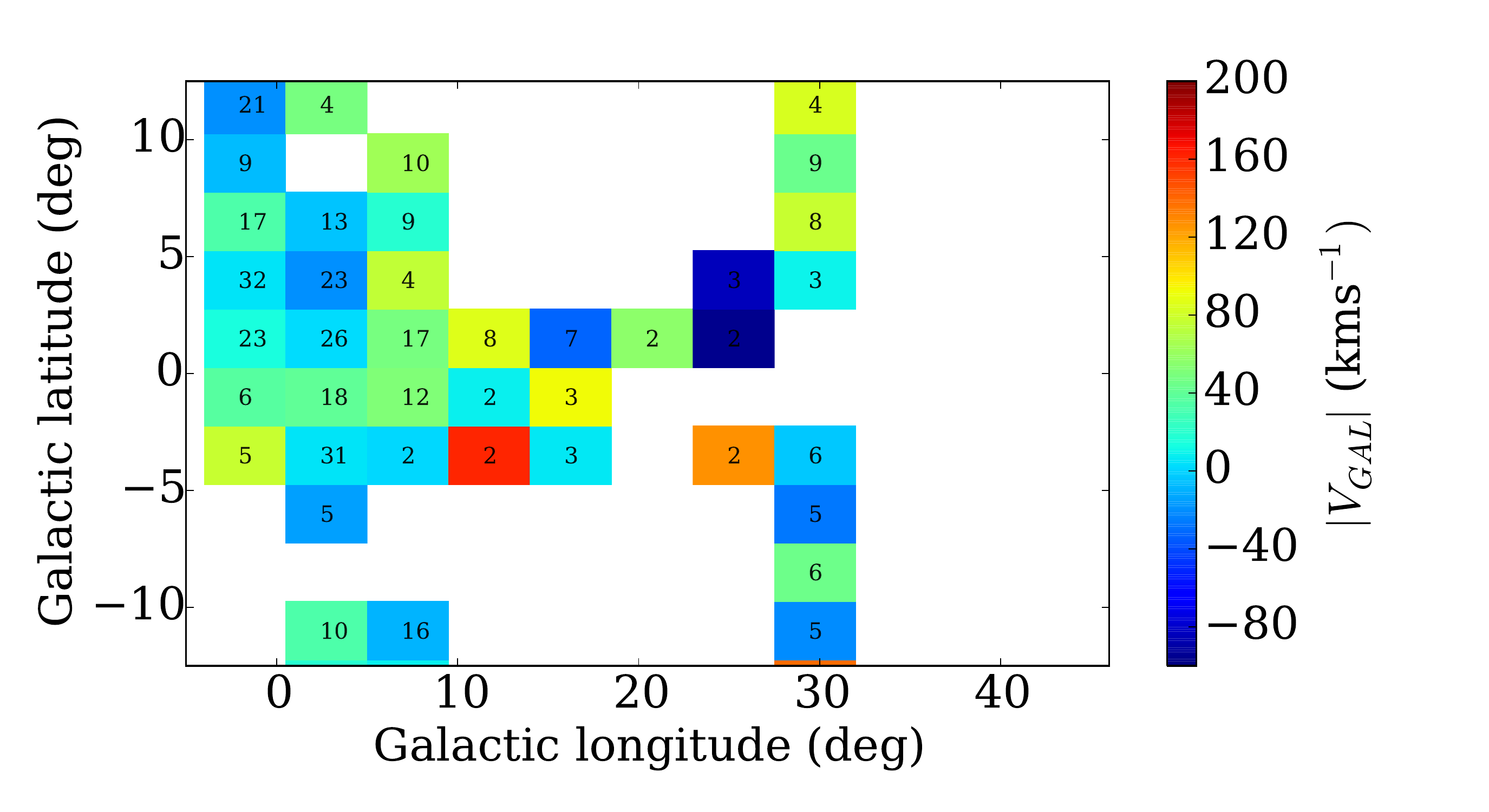}
\caption{D: \feh\ $<$ --1.0} 
\end{subfigure}
  \caption{Mean velocity maps for metallicity populations A, B, C, and D (containing 2700, 2700, 1700 and 450 stars, respectively).  The rotation is cylindrical for populations A,B,C. }
  \label{fig:ABCD}
\end{figure*}

\begin{figure*}
    \begin{subfigure}[b]{0.5\linewidth}
\flushright
    \includegraphics[width=0.9\linewidth]{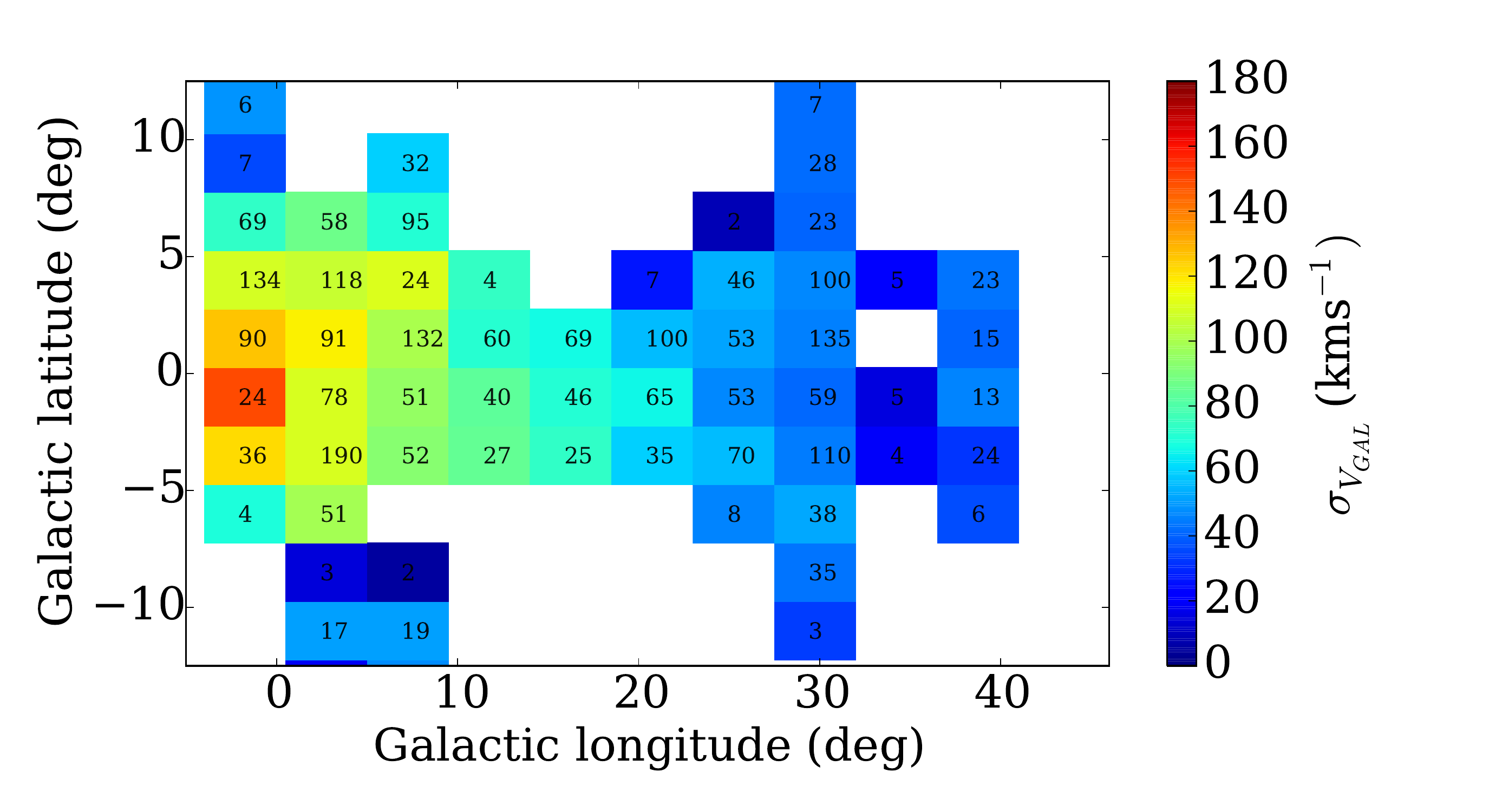} 
\caption{A: \feh\ $>$ 0}
  \end{subfigure}%
  \begin{subfigure}[b]{0.5\linewidth}
\flushright
    \includegraphics[width=0.9\linewidth]{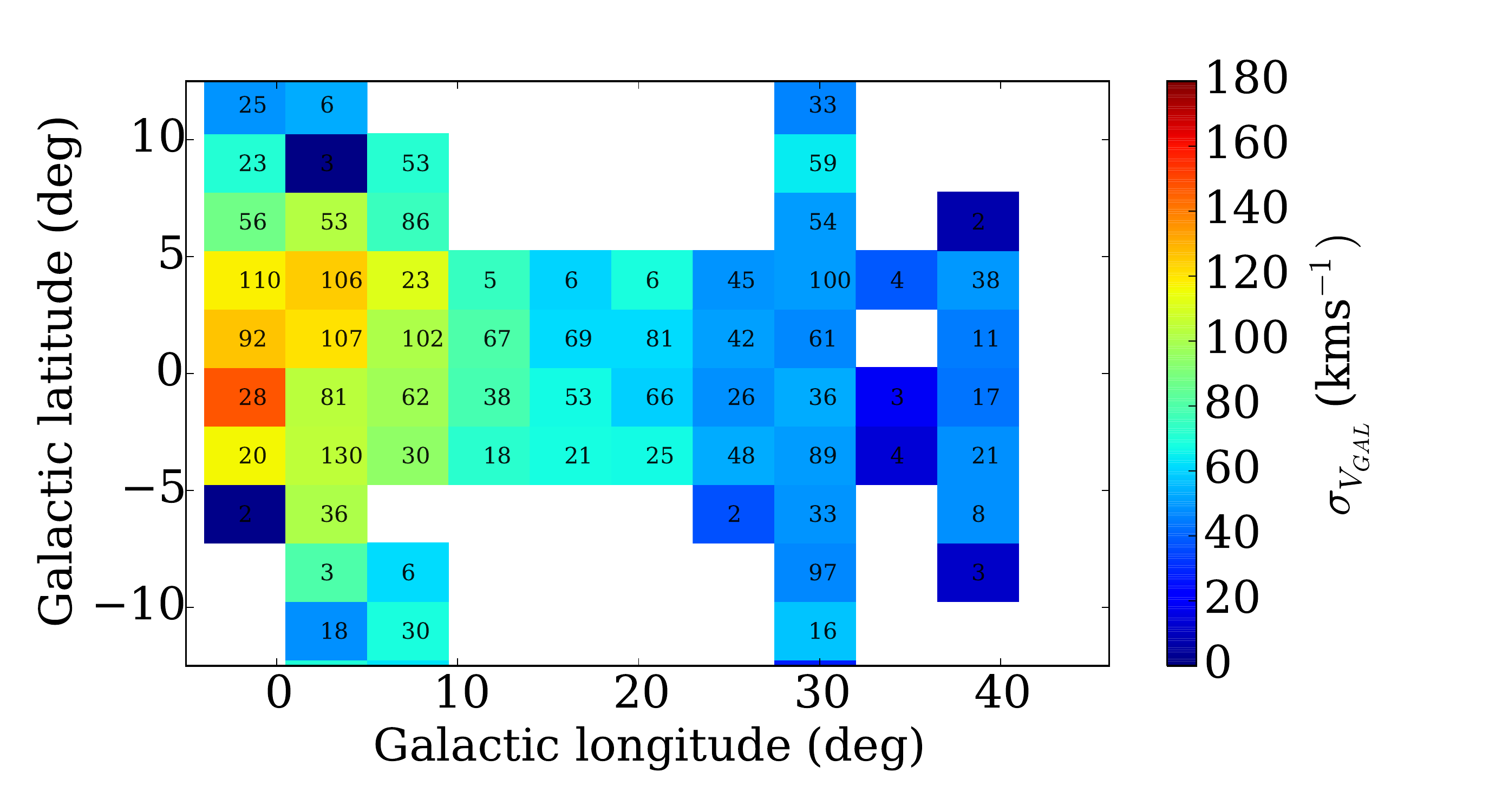} 
\caption{B --0.5 $<$ \feh\ $<$  0 } 
\end{subfigure}
    \begin{subfigure}[b]{0.5\linewidth}
\flushright
    \includegraphics[width=0.9\linewidth]{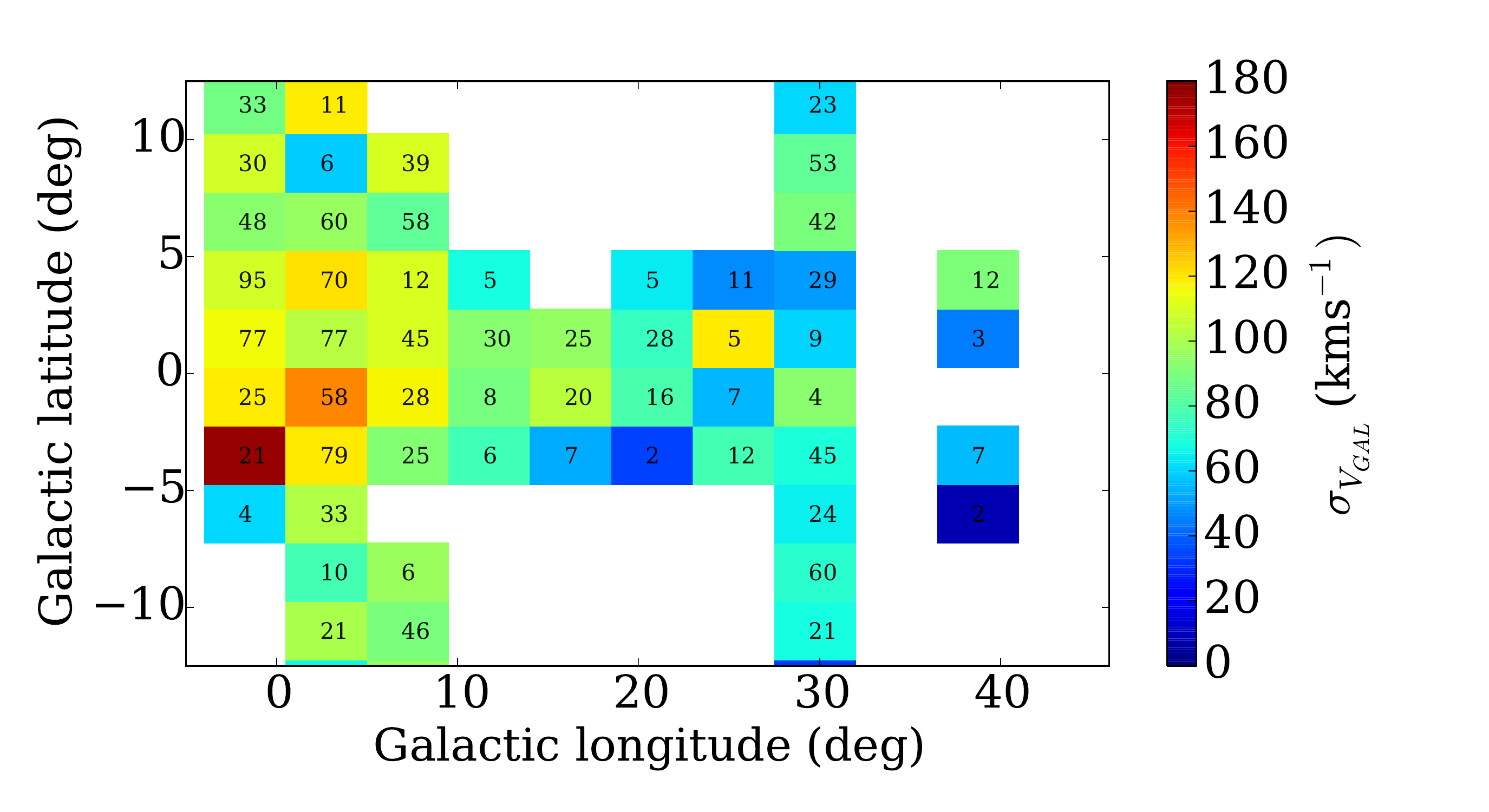}
\caption{C: --1.0 $<$ \feh\ $<$ --0.5}
  \end{subfigure}%
  \begin{subfigure}[b]{0.5\linewidth}
\flushright
    \includegraphics[width=0.9\linewidth]{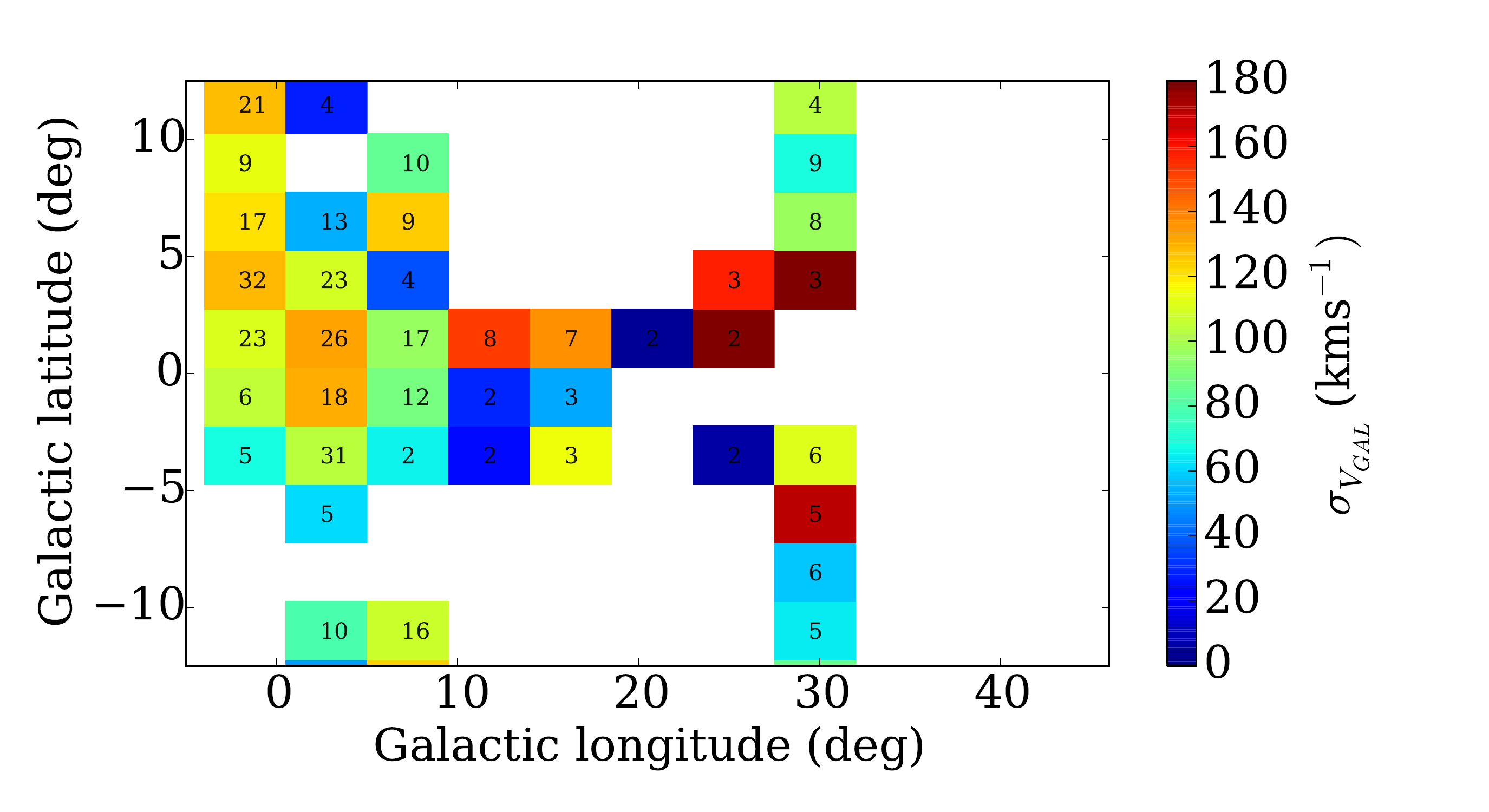} 
\caption{D \feh\ $<$ --1.0} 
\end{subfigure}
    \caption{Velocity dispersion maps of metallicity populations A, B, C, and D.  {The dispersion is similar for stars with \feh\ $>$ --0.5 and flattens {at} higher longitudes and latitudes for the more metal-poor stars.}}
  \label{fig:ABCD2}
\end{figure*}

 Figures~\ref{fig:ABCD} and \ref{fig:ABCD2} present the rotation and dispersion profiles of stars in these four metallicity bins in more detail, across $(l,b)$, where there are 2700, 2700, 1700 and 450 stars in each of these maps, respectively.  

{The \apogee\ mean rotation trends for the stars in the plane ($|b|$ $<$ 2$^\circ$) shown in Figures \ref{fig:ABCD} and \ref{fig:ABCD2} are similar to the higher latitude observations from ARGOS \citep{Ness2013b}. As seen in Figure \ref{fig:ABCD}, the rotation is cylindrical for populations A, B, and C, and increases smoothly with longitude.  Although stars in bin C show a similarly cylindrical rotation profile to stars {with} \feh\ $>$ {--0.5}, they have an overall slower rotation profile than the more metal-rich stars in A and B. The stars in the most metal-poor bin, D, are no longer cylindrically rotating but display a relatively flat rotation profile with longitude across $(l,b)$}.

{Similarly to the bottom panel of Figure \ref{fig:ABCD3}, the velocity dispersion maps of Figure \ref{fig:ABCD2} demonstrate that the stars in the metallicity bins A and B share a similar dispersion profile across $(l,b)$. The dispersion decreases with longitude and flattens for populations A--C. Inside of $l$ $<$ 10$^\circ$, the dispersion decreases as a function of latitude, but is fairly constant with latitude outside of the boxy extent of the bulge.  The most metal-poor stars (population D), display both high and low dispersions, although there are only a few stars in many of these fields. }

\section{Discussion}

The kinematics of the bulge mapped by \apogee\ demonstrate that at latitudes $|b| < 5^\circ$, stars with $\feh\ > -1.0$ have cylindrical rotation, similar to patterns farther from the plane mapped by other surveys (e.g. ARGOS, BRAVA, and GIBS), with a smooth transition in longitude out into the disk. The trends in the rotation as a function of metallicity at $|b| < 5^\circ$ confirm and extend the findings from the ARGOS survey (determined for $|b| > 5^\circ$). The most metal-rich stars, with $\feh\ > -0.5$ (populations A and B in Section~\ref{sec:kinematics_feh}), show the fastest rotation, and the stars with metallicities  $-1.0 < \feh\ < -0.5$ (population C in Section~\ref{sec:kinematics_feh}), although also rotating cylindrically, have the slowest mean rotation among all stars with $\feh\ > -1.0$.  The similar chemodynamical behavior of stars in the mid-plane compared to those at higher latitudes is consistent with bulge formation via instabilities of the Galactic disk and agrees well with simple N-body models of bulge formation from a bar that is formed in the disk. 

In the kinematically hottest region, inside the boxy part of the bulge with $(|l|,|b|) < (5^\circ,5^\circ)$, there appears to be no metallicity gradient in $(l,b)$, as previously reported from smaller samples.  This result is in contrast to higher latitudes, where the metallicity decreases by about --0.45 dex/kpc along the minor axis \citep[e.g.][]{zoccali2008, Ness2013a}. A similar gradient is seen in latitude across the bulge at larger longitudes, and a weaker gradient is present across longitude, for $(|l|,|b|)$ $>$ (5$^\circ$,5$^\circ)$ \citep{Gonzalez2011}. These gradients likely reflect the details of the redistribution of stars into the boxy/peanut bulge from the bulge-forming dynamical instabilities \citep[e.g.,][]{PdiM2015} seen in N-body models.  In these models, stars that are found in the inner bulge region typically originate from the innermost region of the disk, and this kinematically hot inner region today likely reflects a well mixed population built from stars of a limited radial extent. 

There are metal-rich, alpha-poor stars concentrated to within $b$ $<$ 2$^\circ$. These stars are likely part of the bar in the plane beyond the boxy extent of the bulge $l$ $>$ 10$^\circ$ \citep{Wegg2015}; these stars show a smooth chemodynamical transition into the disk. The super-thin bar reported by \citet{Wegg2015} does not have any clear corresponding signature in the chemistry of \apogee\ stars that differentiates them from all stars within $|b|$ $<$ 2$^\circ$. 

The chemodynamics of stars {with} \feh\ $>$ --1.0 suggests a common origin of formation in the disk, although stars with --1.0 $>$ \feh $>$ --0.5 show kinematics that indicates these stars are not part of the boxy structure. The small fraction ($\approx$ 5\%) {of metal-poor bulge stars with \feh\ } $<$ --1.0 show a slower rotation and flatter dispersion that aligns them with a halo population in the inner most region or else a unique metal-poor bulge population formed at early times \citep[see][]{Howes2015, Koch2015}.

The stars that are found at larger $(|l|,|b|)$ in the bulge, in the corners of the boxy/bulge structure, for example, at $(12^\circ,12^\circ)$ are typically redistributed there from further out in the disk and it has been demonstrated that during bulge formation the stars are preferentially redistributed according to their initial phase-space \citep[e.g.][]{Inma2013, PdiM2015}. This redistribution effectively maps the metallicity gradient from the early disk into the bulge. The signatures of this mapping are contained in the relationship between the kinematics and \feh-\alphafe\  of stars as a function of $(R,z)$ today. Therefore, chemodynamical models, or models containing star formation that can track stellar ages, may provide important insights to the interpretation of the trends seen in Figures~\ref{fig:ABCD} and \ref{fig:ABCD2} and discussed in Section 5.3.

\section{Conclusions}

With the low-latitude data from \apogee, covering regions of the bulge never before explored with significant stellar samples, we have mapped the kinematic profile of the bulge in rotation and dispersion out into the disk. We have demonstrated that these profiles are consistent with the chemodynamical findings reported by other surveys at higher latitudes and report new details on the bulge chemodynamics.  We report a population of metal-rich, {alpha-poor} stars that populate the plane region within $b$ $<$ 2$^\circ$, {with metal-rich stars being present at the end of the thin bar, which reaches beyond the boxy profile of the bulge, to $l$ $\approx$ 25--30$^\circ$}.  These metal-rich stars at the end of the bar transition smoothly out into the disk in rotation and dispersion. 

{The kinematic profile of the \apogee\ stars show a dependence on \feh\ that is very similar to the trends seen in ARGOS, with stars with \feh\ $>$ --1.0 having cylindrical rotation and stars with \feh\ $>$ --0.5 showing a dispersion profile characteristic of N-body models of boxy bulges formed from the disk. All stars with \feh\ $>$ --1.0 show a smooth kinematic transition from the bulge region into the disk and likely have an origin in the disk. However, stars with --1.0 $<$ \feh\ $<$ --0.5 do not appear to belong to the boxy bulge morphology {given that they do not show the split red clump feature, which is the signature of the boxy/peanut structure} \citep{Ness2012,Uttenthaler2012}. Furthermore, the dispersion profile of stars with \feh\ $<$ --0.5 does not match the N-body models of boxy peanut bulges. The small fraction of stars (5\%) with \feh\ $<$ -1.0 show kinematics that are dissimilar in both rotation and dispersion to N-body models and the more metal-rich stars.  These metal-poor stars do not transition smoothly in kinematics out into the disk. }

Our kinematic maps for the bulge show remarkable similarity to those of other galaxies with boxy/peanut bulges. This result is perhaps not surprising, given the Milky Way's boxy/peanut bulge profile that has been revealed in detail in the recent work of \citet{Wegg2015} and \citet{Portail2015a} as a typical boxy and off-center X-shaped structure.   The larger data set that will be obtained with \apogee-2 will enable a more detailed and further examination of the chemo-kinematics of the bulge. This includes mapping trends in \alphafe\ and also individual elemental abundances and also testing the symmetry around the major axis and at the ends of the bar, at both positive and negative longitudes.

\section*{Acknowledgements}

We thank the referee for {a} constructive report which has improved the clarity of this work. 

We thank Ortwin Gerhard (MPE), Chris Wegg (MPE) and Matthieu Portail (MPE), Eddie Schlafly (MPIA) and Andy Casey (IoA, Cambridge) for helpful discussions. 
The research has received funding from the European Research Council under the European
Union's Seventh Framework Programme (FP 7) ERC Grant Agreement n.
[321035].

E.A. acknowledges financial support from the
CNES (Centre National d'Etudes Spatiales - France) and from the People
Programme  (Marie Curie Actions) of the European Union's Seventh
Framework Programme FP7/2007-2013/ to the DAGAL network under REA
grant agreement number PITN-GA-2011-289313. 

G.Z. was supported by an NSF Astronomy \& Astrophysics Postdoctoral Fellowship under Award No.\ AST-1203017.

J.A.J. acknowledges support from NSF AST-1211853.

Funding for SDSS-III has been provided by the Alfred P. Sloan Foundation, the Participating Institutions, 
the National Science Foundation, and the U.S. Department of Energy Office of Science. The SDSS-III web site is \url{h!tp://www.sdss3.org/}.

SDSS-III is managed by the Astrophysical Research Consortium for the Participating Institutions of the SDSS-III Collaboration
 including the University of Arizona, the Brazilian Participation Group, Brookhaven National Laboratory, Carnegie Mellon University, 
 University of Florida, the French Participation Group, the German Participation Group, Harvard University, the Instituto de Astrofisica 
 de Canarias, the Michigan State/Notre Dame/JINA Participation Group, Johns Hopkins University, Lawrence Berkeley National Laboratory, 
 Max Planck Institute for Astrophysics, Max Planck Institute for Extraterrestrial Physics, New Mexico State University, New York University, 
 Ohio State University, Pennsylvania State University, University of Portsmouth, Princeton University, the Spanish Participation Group, 
 University of Tokyo, University of Utah, Vanderbilt University, University of Virginia, University of Washington, and Yale University

%\bibliography{ak.bib}

\section*{Appendix}

\begin{table*}[h!]
\caption{Partial column excerpt from the online Table of four stellar labels (\teff, \logg, \feh, \alphafe\ ) determined by \tc\ for  APOGEE DR12 stars toward the bulge and inner disk with their $\chi_{reduced}^2$ values and distances.  \apogee\ targeting flags should be used in the selection of stars and main sequence stars are provided to identify these stars as such, but stellar parameters are unreliable for these stars which should be excluded from analysis. } 
\begin{tabular}{| c | c | c |  c | c | c |  c | c | } 
\hline
\small{star ID}  & \teff\ & \logg\ & \feh\ & \alphafe\  & $\chi_{reduced}^2$ & distance & \vgal\  \\
\small{(2MASS)} & K & dex &  dex  & dex &&  kpc  & \kms\  \\    
\hline
2M18350759-0508445 & 3727.8 & 0.7 & -0.18 & -0.0 & 1.5 & 12.1 & 159.1 \\
2M18354096-0452009 & 3952.6 & 1.4 & 0.2 & -0.03 & 1.4 & 5.5 & 130.1 \\
2M18354152-0533504 & 3952.8 & 0.7 & -0.34 & 0.04 & 3.1 & 9.7 & 157.4 \\
2M18354345-0453021 & 4636.0 & 2.6 & 0.37 & -0.0 & 1.2 & 2.9 & 120.8 \\
2M18354380-0425029 & 4053.5 & 1.4 & -0.44 & 0.23 & 1.2 & 10.6 & 205.7 \\
2M18355233-0501398 & 4604.9 & 2.3 & 0.35 & 0.01 & 0.9 & 1.1 & 93.8 \\
2M18361149-0608046 & 4418.3 & 1.9 & 0.34 & 0.02 & 1.3 & 1.6 & 128.6 \\
2M18361409-0456347 & 4159.6 & 1.8 & 0.52 & -0.07 & 2.6 & 3.9 & 141.7 \\
2M18362011-0510251 & 4432.5 & 2.1 & 0.21 & 0.03 & 3.5 & 3.8 & 145.8 \\
2M18362181-0545341 & 4625.9 & 2.5 & -0.36 & 0.21 & 2.0 & 2.5 & 131.7 \\
\hline
\end{tabular}
\label{tab:online} 
\end{table*}
\end{document}